\documentclass[manuscript,11pt]{aastex}
\pdfoutput=1

\usepackage[pdftex]{color}
\usepackage{ifthen}
\usepackage{natbib}
\definecolor{MyChange}{rgb}{0,0,0}
\definecolor{MyChangeB}{rgb}{0,0,0}

\newboolean{movFull}
\setboolean{movFull}{false} 
\ifthenelse {\boolean{movFull}}     {\usepackage{animate}}{}

\def\uset{\{{\bf U}\}}

\def\gapprox{\hbox{ \lower0.5ex\hbox{$\sim$} \kern-1.1em 
                 \raise0.5ex\hbox{$>$} }} 
\def\lapprox{\hbox{ \lower0.5ex\hbox{$\sim$} \kern-1.1em 
                 \raise0.5ex\hbox{$<$} }} 

\lefthead{H. Levison et al.}
\righthead{LIPAD}

\begin{document}

\title{A Lagrangian Integrator for Planetary Accretion and Dynamics
 (LIPAD)}

\author{Harold F$.$ Levison}
\affil{Department of Space Studies, Southwest Research Institute,
       Boulder, CO, USA 80302}
\authoremail{hal@boulder.swri.edu}

\author{Martin J$.$ Duncan}
\affil{Department of Physics, Engineering Physics, \& Astronomy\\ Queen's University\\
Kingston, Ontario, K7L 3N6\\ Canada}

\and

\author{Edward Thommes}
\affil{Department of Physics\\ University of Guelph\\ Guelph, Ontario,
N1G 2W1\\ Canada} 

\received{February 15, 2011}

\accepted{}

\newpage

\begin{abstract}

  We presented the first particle based, Lagrangian code that can
  follow the collisional/accretional/dynamical evolution of a large
  number of km-sized planetesimals through the entire growth process
  to become planets.  We refer to it as the {\it Lagrangian Integrator
    for Planetary Accretion and Dynamics} or {\it LIPAD}.  LIPAD is
  built on top of SyMBA, which is a symplectic $N$-body integrator
  \citep{{Duncan:1998aj116:2067}}.  In order to handle the very large
  number of planetesimals required by planet formation simulations, we
  introduce the concept of a {\it tracer} particle.  Each tracer is
  intended to represent a large number of disk particles on roughly
  the same orbit and size as one another, and is characterized by
  three numbers: the physical radius, the bulk density, and the total
  mass of the disk particles represented by the tracer.  We developed
  statistical algorithms that follow the dynamical and collisional
  evolution of the tracers due to the presence of one another.  The
  tracers mainly dynamically interact with the larger objects ({\it
    planetary embryos)} in the normal $N$-body way.  LIPAD's greatest
  strength is that it can accurately model the wholesale
  redistribution of planetesimals due to gravitational interaction
  with the embryos, which has recently been shown to significantly
  affect the growth rate of planetary embryos
  \citep{Levison:2010AJ:139:1297}. We verify the code via a
  comprehensive set of tests which compare our results with those of
  Eulerian and/or direct $N$-body codes.

\end{abstract}


\received{}

\accepted{}

\newpage

\section{Introduction}
\label{sec:intro}

The construction of a comprehensive, end-to-end model of the
accumulation of the terrestrial planets and giant planet cores has
been an elusive goal for planetary scientists because of the huge
dynamic range inherent in the problem.  The region of the
proto-planetary disk from which the planets formed originally
contained something like $10^{\sim\!14}$ objects with radii
{\color{MyChange}perhaps as small as $\sim\!100\,$m
  \citep{Weidenschilling:2011Icar:214:671} or as large as $1000\,$km
  \citep{Johansen:2011EM&P:108:39} depending on the planetesimals
  formation model}.  These objects grew into the planets via a process
that includes both complex collisional (both accumulation and
fragmentation) and dynamical evolution. In addition, at different
stages of this process, the action occurred on very different temporal
and physical scales, making the construction of comprehensive models
very difficult.

Take, for an example, the formation of terrestrial planets.  Studies
have shown that {\color{MyChange} once the first macroscopic
  planetesimals have formed (which is a field of study in itself),}
solid body growth can occur in three distinct stages.  In the first
stage, planetesimals grow by so-called {\it runaway} accretion
\citep{Wetherill:1989Icar:77:330, Greenberg:1978Icar:35:1}.  During
this stage, the largest objects do not affect the dynamical state of
the rest of the disk and so an object's mass accretion rate scales as
$M^{4/3}$.  As a result, the largest bodies grow the fastest ---
mainly by feeding off of much smaller objects.
\cite{Ida:1993Icar:106:210} showed that runaway accretion ends when
the growing planets are only roughly $100\times$ their original mass.
Because this stage requires the study of hundreds of billions of
objects, the codes used to study it employ Eulerian statistical
algorithms which divide the problem into a multidimensional grid,
usually 2-dimensional in heliocentric distance and size
\citep{Wetherill:1989Icar:77:330, Spaute:1991Icar:92:147,
  Kenyon:1999aj118:1101, Kenyon:2001aj:121:538,
  Morbidelli:2009Icar:204:558, Bromley:2011apJ:731:101} which evolves
the total mass, and RMS eccentricity and inclination in each bin.
These codes usually accurately follow the detailed
collisional/fragmentational evolution of system, while using
relatively simple, semi-analytic equations to evolve the dynamics.
These dynamical equations are appropriate in this stage because the
dynamics are local and well behaved --- there is little dynamical
mixing and the surface density of the system remains smooth.

In the middle stage, the largest bodies become big enough to
gravitationally ``stir their own soup'' of planetesimals
\citep{Ida:1993Icar:106:210, Kokubo:1998Icar:131:172,
  Kokubo:2000Icar:143:15, Thommes:2003Icar:161:431,
  Chambers:2006ApJ:652L:133}, and thus the mass accretion rate of the
largest bodies scales as $M^{2/3}$.  In this phase, the largest few
objects at any given time are of comparable mass.  As the system
evolves, the mass of the system is concentrated into an
ever-decreasing number of bodies, known as {\it planetary embryos}, of
increasing masses and separations.  This stage ends at a given
location in the disk when the surface density of the local
``oligarchs'' becomes similar to that of the planetesimals
{\color{MyChange}\citep{Kenyon:2006aj:131:1837}}.  This occurs when
the largest bodies reach roughly half their so-called {\it isolation
  mass}, which is the mass they would have if they had consumed all
planetesimals within their gravitational reach.  In the terrestrial
planet region, typical disk models produce isolation masses of only
about Mars mass --- thus a third, very violent, phase must take place
in which these bodies' orbits cross and they collide to form Earth-
and Venus-mass bodies. The middle- and late-stages have mainly been
studied with direct $N$-body simulations \citep[][for
example]{Chambers:1998Icar:136:304, Agnor:1999Icar:172:219,
  Chambers:2001Icar:152:205, OBrien:2006Icar:184:39,
  Raymond:2009Icar:203:644}.  The $N$-body codes accurately follow the
dynamical evolution of the system, which is necessary because there is
much mixing and chaotic behavior.  Studies of these stages are
required to represent the large number of planetesimals remaining in
the system by a smaller number of more massive {\it tracer} particles
in order to make the problem computationally tractable.  In addition,
they assume that when two bodies collide, they merge with 100\%
efficiency; there is no fragmentation.

There have been a couple of attempts at constructing an end-to-end
simulation of planet formation that started with a population of small
planetesimals and built a complete planetary system \citep[][for
example]{Spaute:1991Icar:92:147, Weidenschilling:1997Icar:128:429,
  Kenyon:2006aj:131:1837, Bromley:2011apJ:731:101}.  These have
employed codes that graft an $N$-body algorithm onto Eulerian
statistical code.  The dynamics of the growing planetary embryos are
handled correctly by the $N$-body algorithm, the
accretion/fragmentation of the planetesimals are handled by the
Eulerian code, and the interaction between the two populations are
handled via analytical expressions (for example, applying dynamical
friction to the embryos by the planetesimals).  The embryos can affect
the eccentricities and inclinations of the planetesimals, but not
their surface density distribution.

The last point above is likely to be a serious limitation of these
algorithms.  In \citet[][hereafter LTD10]{Levison:2010AJ:139:1297} we
showed that the growth rate of planetary embryos is strongly effected
by the wholesale redistribution of planetesimals due to gravitational
interaction with the embryos, themselves.  In particular, we found
that growth can stop if a gap opens around a embryo.  In addition, the
embryos can migrate as a result of gravitational scattering of the
nearby planetesimals \citep[see also][]{Fernandez:1984Icar:58:109,
  Hahn:1999AJ:117:3041, Ida:2000AJ:534:428, Levison:2007prpl.conf:669,
  Kirsh:2009Icar:199:197}.  This so-called {\it planetesimal driven
  migration} can significantly enhance growth
\citep[LTD10;][]{Minton:2012Icar:sub}.  Unfortunately, this result
calls into question the bulk of the models of the early stages of
planet formation because they rely on algorithms that do not take this
process into account.

We realized that in order to adequately incorporate our results into
full planet formation simulations would require a totally new,
Lagrangian approach to the problem.  Fortunately, we also realized
that the code used in LTD10 supplied us with a basic structure in
which to develop this algorithm.  Here we report on the first
particle-based Lagrangian code that can follow the
dynamical/collisional/accretional evolution of a large number of
km-sized planetesimals through the entire growth process to become
planets.  We call this code {\it LIPAD} for Lagrangian Integrator for
Planetary Accretion and Dynamics.  In \S{\ref{sec:code}}, we describe
the code in detail.  In \S{\ref{sec:tests}}, we present a
comprehensive set of tests and show that LIPAD represents the behavior
of a systems containing a large number of planetesimals better than
Eulerian codes.  Finally, our conclusions are presented in
\S{\ref{sec:conl}}.

\section{LIPAD}
\label{sec:code}

LIPAD is built on top of our $N$-body code known as SyMBA
\citep{Duncan:1998aj116:2067} and it is an extension of the code used
to study giant planet core formation in LTD10. SyMBA is a symplectic
algorithm that has the desirable properties of the sophisticated and
highly efficient numerical algorithm known as the Wisdom-Holman Map
\citep[WHM,][]{Wisdom:1991AJ:102:1528} and that, in addition, can
handle close encounters \citep{Duncan:1998aj116:2067} This technique
is based on a variant of the standard WHM, but it handles close
encounters by employing a multiple time step technique introduced by
\cite{Skeel:1994ANM:1:191}. When bodies are well separated, the
algorithm has the speed of the WHM method.  However, whenever two
bodies suffer a mutual encounter, the time step for the relevant
bodies is recursively subdivided in a way that keeps the system
symplectic.

Since we cannot possibly follow the evolution of $10^{\sim\!14}$
bodies, we introduce four classes of particles to LIPAD: two types of
embryos, to represent individual large objects (these will be
described below), and two types of {\it tracers}, to represent the
small-size end of the population.  Each tracer is intended to
represent a large number of comparably-sized planetesimals on roughly
the same orbit.  Each tracer will be characterized by three numbers:
the physical radius $s$ and the bulk density $\rho$ of its constituent
planetesimals, as well as the total mass of the particles it
represents, $m_{\rm tr}$.  As the system evolves, $m_{\rm tr}$ and
$\rho$ remain fixed, while, as we describe in detail below, the
tracer's orbit and $s$ change.  As a result, the number of
planetesimals that the tracer represents, $N_{\rm tr}\!=\!m_{\rm
  tr}/{{4\over 3}\pi\rho s^3}$, also changes.  It is important to note
that in order to strictly conserve mass during a simulation $N_{\rm
  tr}$ is a real number and not an integer.  Although this might seem
odd, it is a reasonable approach because the combined tracer
population in intended to statistically represent a distribution of
the much larger number of planetesimals.

\begin{figure}[h!]
\includegraphics[scale=0.45]{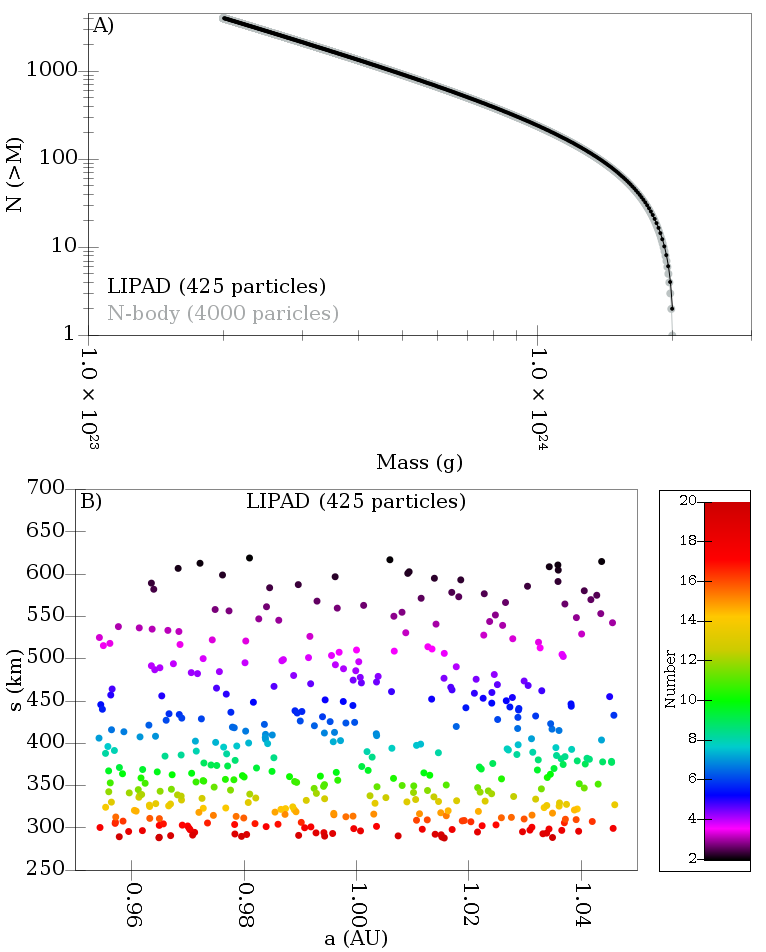}
\caption{\footnotesize \label{fig:KIic} An example of how a population
  of planetesimals is distributed among tracers.  In particular, 425
  tracers, each of mass $4.2 \times 10^{24}\,$g, are representing 4000
  real objects with masses between $10^{23}$ and $10^{24}$ gm.  The
  planetesimal differential mass distribution is a power law with
  $N(s)\,ds \propto s^{-2.6}$.  A) The cumulative number distribution.
  The gray curve shows the distribution of the 4000 objects used in an
  $N$-body calculation described in \S{\ref{ssec:KI}}.  The black
  curve shows the tracers.  B) Each dot represents an individual
  tracer with radius $s$ as a function of semi-major axis ($a$).  The
  color indicates the number of planetesimals that the tracer
  represents ($N_{\rm tr}$).}
\end{figure}

Perhaps the best way to illustrates how tracers operate is with a
simple example.  In \S{\ref{sec:tests}}, we describe a test of LIPAD
where we attempt to reproduce the terrestrial planet accretion
calculation of \citet{Kokubo:2000Icar:143:15}, who studied a disk that
extended from 0.99 to 1.01 AU and contained $0.3\,M_\oplus$ of
material. One of their runs contained 4000 particles with initial
masses of between $10^{23}$ and $10^{24}$ gm; distributed so that
$N(s)\,ds$ is a power-law with a slope of -2.6 (see
Figure~\ref{fig:KIic}A). They followed the system with a full N-body
code.  We performed the same simulation with LIPAD, using 425 tracers.
While the results of this calculation are discussed in the section
describing the tests we performed on LIPAD (\S{\ref{ssec:KI}}), a look
at the initial conditions give us an opportunity to clarify how
tracers work.

Figure~\ref{fig:KIic}B shows the initial particle distribution of our
calculation.  In particular, we plot $s$ as a function of semi-major
axis, $a$.  The tracers each have the same total mass, and so the
color represents the total number of planetesimals that each tracer
represents.  In this case as $s$ decreases from 620 to $290\,$km,
$N_{\rm tr}$ increases from 2 to 20, but $N_{\rm tr}$ can, in
principal, become much larger (see some examples in
\S{\ref{sec:tests}}). Thus, it might be better to think of each tracer
as representing a clump of material of mass $m_{\rm tr}$ than as an
individual object in an $N$-body simulation.

The tracers dynamically interact with the larger objects (the two
classes of embryos, see below) in the normal $N$-body way.  The
innovative aspect of this code is how the tracers interact with each
other.  In particular, we employ Monte Carlo algorithms to evolve the
sizes and random velocities of a tracer's constituent planetesimals
based on the location and behavior of the neighboring tracers.  As
described in more detail below, we employ statistical algorithms to
incorporate dynamical friction, viscous stirring, collisional damping,
and accretion/fragmentation. The code is designed so that as objects
grow (i.e$.$ $s$ increases and $N$ decreases) a tracer can be {\it
  promoted} to an embryo if $N$ becomes equal to 1.  When an object is
promoted it becomes an embryo with mass $m_{\rm tr}$.

Embryos and tracers interact with one another through the SyMBA
$N$-body routines.  This leads to a problem when a tracer is promoted.
Before promotion, the object in general sees itself embedded is a sea
of much smaller objects through the Monte Carlo routines.  If we were
to simply promote an object to an embryo, it would suddenly see itself
in a swarm of like-mass objects because the other tracers have a total
mass of $m_{\rm tr}$, as well.  

Embryos and tracers interact with one another through the SyMBA
$N$-body routines.  This leads to a problem when a tracer is promoted.
Before promotion, the object in general sees itself embedded
{\color{MyChange} in a sea of much smaller planetesimals.  This is due
  to the fact that, although these small planetesimals are represented
  by much more massive tracers, their effects are felt through the
  Monte Carlo routines.  If we were to simply promote an object to an
  embryo, we would have a problem because the embryos directly
  interact with the tracers and immediately after promotion the embryo
  is only slightly more massive than they are.  In this case, the
  perturbations from tracers are much too noisy due to their coarse
  mass resolution.  To avoid these} unphysically large gravitational
scatterings, we introduce the concept of a {\it sub-embryo}.
Sub-embryos can respond to analytically computed dynamical friction
and planet migration effects of the planetesimals (see below), and can
collide with them, while interacting with the rest of the embryos
through the SyMBA $N$-body routines.  A user is free to set the mass
at which a sub-embryo becomes a {\it full-embryo}.  Based on the work
on planetesimal-driven migration of \citet{Kirsh:2009Icar:199:197}, we
recommend that this boundary be at least $100~m_{\rm tr}$.

Finally, LIPAD is designed to be able to handle a collisional cascade
that can potentially remove material from the system.  This is
accomplished by defining a class of tracer that represents a
population of very small objects that we call {\it dust}, for lack of
a better word.  If the size of a tracer evolves so that $s < s_{\rm
  min}$, it is demoted to {\it dust} with a size $s_{\rm dust}$.  Both
$s_{\rm dust}$ and $s_{\rm min}$ are free parameters of the code. The
user has several options concerning the behavior of the dust.  One is
for dust particles to behave like a tracers with $s = s_{\rm min}$ in
every respect except that they can no longer fragment.  The user also
has the option so that dust tracers do not interact with other tracers
at all, but fully interact with the embryos and sub-embryos via the
$N$-body routines.  In addition, users can, if they wish, apply a
fictitious force to these particles that represents the
Poynting-Robertson drag \citep{Robertson:1937MNRAS:97:423} of a
particle with $s_{\rm dust}$.  This allows the particles to slowly
drift through the system, eventually being removed by hitting the Sun
(if they do not get accreted by an embryo). {\color{MyChange} Note
  that we disable the aerodynamic drag terms in the equation of motion
  of a dust particle.}

In summary, we employ four classes of particles:

\begin{enumerate}
\item{} {\it Full-Embryos:} These objects interact with all classes of
  particles through the normal $N$-body routines, i.e$.$ through the
  direct summation of individual forces.  The $N$-body routines also
  monitor whether physical collisions occur.  The algorithm that LIPAD
  uses to handle these collisions is described in
  \S{\ref{sssec:EmCol}}.
\item{} {\it Sub-Embryos:} These objects interact with full-embryos
  and each other through the $N$-body routines.  However, the only
  dynamical effect that the tracers have on them is through analytic
  dynamical friction and planet migration routines
  (\S{\ref{ssec:embryos}}).  Collisions are handled in the same way as
  those of the full-embryos.
\item{} {\it Tracers:} These objects gravitationally interact with
  each other through Monte Carlo routines that include viscous
  stirring, dynamical fraction, and collisional damping.  They
  gravitationally interact with the embryos via the $N$-body
  routines. During a close encounter with a sub-embryo, however, the
  mass of the tracer is set to zero so that the orbit of the
  sub-embryo is not perturbed.
\item{} {\it Dust Tracers:} These are tracers that can no longer
  fragment.  The user can set the code so that these objects do not
  interact with the other tracers.  However, they always interact with
  the embryos via the $N$-body routines.  The user also has the option
  to apply Poynting-Robertson drag.
\end{enumerate}

Note that since embryos and tracers interact with one another through
the $N$-body routines, LIPAD accurately handles the redistribution of
the planetesimals and planet migration.  We now describe each of these
classes in detail.  We start with the tracers because they are the
most complex.

\subsection{Behavior of the Tracers}
\label{ssec:tracer}

Both the dynamical and collisional evolution of the tracers due to the
presence of other tracers are handled through statistical algorithms
that change the orbit of the tracer and the value of its radius $s$.
In order to perform these calculations, we first need to determine the
encounter rate between planetesimals in our disk.  This is followed by
a series of calculations of the response of the tracer to these
encounters.

\subsubsection{Tracer-Tracer Encounter Rates}
\label{sssec:rates}

{\color{MyChange} The first step in our calculation is to determine
  the probability, $p$, that a particular planetesimal will suffer an
  encounter with another planetesimal during a timestep, $dt$}.  In
what follows we refer to this object as the {\it target} and the
potential encounter partner as the {\it interloper}.  We perform this
calculation for two types of encounters: physical collisions, $p_{\rm
  col}$, and gravitational scattering events, $p_{\rm grav}$.  We
employ the particle-in-a-box approximation.  In particular, $p \sim
n\, \sigma\, w\, dt$, where $n$ is the local number density of the
disk particles, $\sigma$ is the cross-section of the encounter,
{\color{MyChange}and} $w$ is the mean encounter
velocity{\color{MyChange}.  In order for these routines to seamlessly
interface with the $N$-body algorithm, the code requires that the
statistical timestep be an integer multiple of the $N$-body timestep.
This integer is an input parameter to the code.  It is also} important
to note that $n$ is not the number density of tracers, but the total
number density of their constituent planetesimals.

{\color{MyChange} Before we discuss how we calculate the $p$'s, it is
  instructive to clarify how they are used within the code.  As we
  said above, $p$ is the probability that an individual planetesimal
  will suffer an enounter (be it a scattering event or collision) in
  time $dt$.  Another way of thinking about $p$ is that it is the
  fraction of planetesimals that suffer this enounter.  Thus, if there
  are $n$ planetesimals, $np$ of them will suffer the encounter.  We
  can, of course, look at it in terms of mass.  If there is a total
  mass of $m$ in these planetesimals, then $mp$ of that suffer the
  encounter. Now, if these planetesimals are presented by $n_t$
  tracers of mass $m_t$, then the number of tracers that suffer the
  encounter is $mp/m_t = p \times m/m_t = p n_t$. The number of
  tracers undergoing the encounter is also $p n_t$.  Thus, we can
  apply $p$ directly to the individual tracers. This leads to a
  different interpretation for the tracers.  Above we defined a tracer
  as `a large number of disk particles on roughly the same orbit and
  size as one another'.  For our purpose here, it is just as valid is
  to think of a tracer as an individual planetesimal that we are
  highlighting and following to illustrate the behavior of the system
  as a whole. That is, they are planetesimals that {\it trace} the
  behavior of the system. In either case, we act on the tracer based
  on the value of $p$ alone.}
 
We expect that $n$, $\sigma$, and $w$ will not only be a function of
time and location in the disk, but of the size of the particles as
well.  After all, in the particle swarm, the equilibrium eccentricity
and inclination of a planetesimals will be a function of its size.  As
a result, in order to accurately determine the $p$'s, we must
integrate $n\, \sigma\, w\,$ over the sizes of the planetesimals at
different locations in the disk.  To accomplish this we divide the
planetesimals into a two-dimensional grid in heliocentric distance,
$a$, and particle size.  The Solar System is first divided into a
series of logarithmically spaced annular rings that, in the
simulations performed here, stretched from $0.5\,$AU to $60\,$AU.  In
all, we divided space into {\color{MyChange} $N_{\rm ring}$} such
rings.  In addition, $s$ is crudely divided into {\color{MyChange}
  $N_{\rm s-bin}$} logarithmically spaced bins.  The range of these
bins {\color{MyChange}, $N_{\rm ring}$, and $N_{\rm s-bin}$} are free
parameters of the code {\color{MyChange}(although unless otherwise
  stated $N_{\rm s-bin}=10$ and $N_{\rm ring}\!=\!1000$ in the tests
  presented in \S{\ref{sec:tests}})}.  We can justify the crude
spacing by noting that this grid representation of the state of the
disk is only used to determine the encounter rates and {\it not} what
happens during the encounter.

As the simulation progresses, we keep track of the tracer particles
moving through each bin and from this calculate: 1) the total number
of planetesimals in that ring, $N_t(i_a,j_s)$, where $i_a$ and $j_s$
are indexes for the heliocentric and size bins, respectively, 2) the
total mass in the bin $M_t(i_a,j_s)$, 3) the average mass of the
planetesimals in the bin, $m_p(i_a,j_s) \equiv \langle {4\over
  3}\pi\rho s^3 \rangle$, and 4) the cylindrical radial and vertical
velocity dispersion of the disk particles, $u_a(i_a,j_s)$ and
$u_z(i_a,j_s)$, respectively.  These numbers are used in the
particle-in-the-box calculation of the encounter probabilities.  

The encounters themselves use the positions, velocities, and sizes of
real objects in the simulation which are chosen at random at the time
of the encounter.  However, we were concerned that there may not be
enough particles in the simulation to find a neighbor close enough to
the target to accurately calculate the result of the encounter.  In
order to increase the pool of potential interlopers, we keep a running
list of a particle's position and velocities as they pass through each
individual ring.  Entries are dropped from this list if they are older
than a parameter $\tau_{\rm update}$.  So, at any time during the
simulation, we have a list of potential interlopers for each
$(i_a,j_s)$ combination.

We allow each tracer to interact with each $s$ bin separately because
we expect that the $p$'s and the results of the encounter to be a
strong function of $s$. First, we determine the heliocentric bin the
object is in, $i_a$.  For each size bin, we calculate the local number
density of planetesimals, $n(j_s)$, from the binned parameters of the
disk.  In particular, the midplane number density is assumed to be
\begin{equation}
  n_0(i_a,j_s) = \frac{N_t(i_a,j_s)}{\sqrt{2} h a},
\end{equation}
where $h= u_z(i_a,j_s) \Omega $ is the scale height of the disk, and
$\Omega$ is the orbital frequency at the particle's position.
Following \citet{Lissauer:1993prpl.conf:1071}, we assume that
\begin{equation}
 \label{Eq:njs}
  n(j_s) = n_0(i_a,j_s) e^{-\frac{\vert z \vert}{h}}.
\end{equation}

Unfortunately, we found through painful experimentation that $n\,
\langle\sigma\, w \rangle$ is not equal to $ n\,
\langle\sigma\rangle\, \langle w \rangle$ because $w$ and $\sigma$ are
correlated with one another when gravitational focusing is taken into
account.  We therefore need to calculate the combined average
$\langle\sigma w \rangle$.  This is done by choosing 10 objects at
random from our running list of potential interlopers calculating
$\sigma$ and $w$ for each of these, and taking the average of the
resulting product.  {\color{MyChange} In systems that are dynamically
  cold, this procedure must take into account that fact that not only
  can relative velocity bring two particles together, but Kepler
  shear as well.  As a result, we take
\begin{equation}
 \label{Eq:wshr}
  w = \sqrt{ (\Delta v_\varpi)^2 + (\Delta v_z)^2 + (\Delta
    v_\phi+v_{\rm shr})^2}, 
\end{equation}
where the $\Delta v$'s are the components of the relative velocity in
cylindrical coordinates, and $v_{\rm shr}$ is the shear velocity taken
over a radius of $\sqrt{\sigma/\pi}$.}

As we already mentioned, we need to calculate two $\sigma$'s: one for
physical collisions and one for gravitational scatterings. The
collision cross-section is simply $\pi(s_{\rm targ}\!+\!s_{\rm
  int})^2F_g$, where $F_g$ is the gravitational focusing factor, and
$s_{\rm targ}$ and $s_{\rm int}$ refer to the size of target and
interloper, respectively.  We employ the formalism of
\cite{Greenzweig:1990Icar:87:40} to calculate $F_g$.

The situation is a bit more complicated for the gravitational
scattering case.  Recall that the cross-section represents an area in
which encounters take place in a plane that contains the target and is
perpendicular to the encounter velocity vector (i.e$.$ the so-called
{\it encounter plane}).  For physical collisions without gravitational
focusing, this area is simply the sum of the objects' physical radii.
In LIPAD, we only include encounters where the interloper gets within
a mutual Hill sphere radius of the target, $r_{\rm H} \equiv a
\left[(m_{targ}\!+\!m_{int})/(3M_\odot)\right]^{1/3}$.  However, care
must be taken when the system is dynamically cold.  If $u_z$ is large
enough that the scale height of the planetesimal disk exceeds the
radius of the Hill's sphere, the disk behaves as if it is fully three
dimensional and encounters are possible anywhere within a circle with
a radius $r_{\rm H}$.  In this case, which is referred to as the {\it
  dispersion-dominated} regime, $\sigma = \pi r_{\rm H}^2$.  However, if
the scale height of the disk, $h$,  is smaller than $r_{\rm H}$, which is
known as the {\it shear-dominated regime}, then encounters can only
occur in the union of $\pi r{\rm _H}^2$ and a slab of height $2h$.  In
this case,
\begin{equation}
\label{Eq:sigg}
\sigma_{\rm grav} = 2r_{\rm H}^2 \sin^{-1}{\left(\frac{h_e}{r_{\rm H}}\right)} + 2h\sqrt{r_{\rm H}^2-h_e^2},
\end{equation}
where $h_e$ is the height of the slab projected into the encounter
plane.  We set the slab height to $2h$ to allow for the fact that the
target can lie above the plane, while the interloper is below, or {\it
  vica-versa}.

With the two $\sigma$'s in hand we calculate $p_{\rm col}$ and
$p_{\rm grav}$.  These are then compared to two random numbers chosen
from a uniform distribution between 0 and 1. (For the remainder of
this paper, we refer to this uniform distribution as \uset.)  If
$p_{\rm col}$ is less than its random number we say that a physical
collision occurs, and if $p_{\rm grav}$ is less than its random
number, then a scattering event occurs.  We now describe how the
tracer responds to these events.

\subsubsection{Collisional Evolution of the Tracers}
\label{sssec:col}

Physical collisions have two effects on the tracers: they damp their
random motions and they cause the sizes of their constituent
planetesimals to change.  In order to avoid double counting, tracers
are only allowed to collide with objects smaller than themselves (this
limitation is also applied when calculating $p_{\rm col}$).

Once we determine that a tracer has suffered a collision with another
disk particle using procedures described in the last section, we need
to determine the characteristics of the impactor.  As we stated above,
as the tracers orbit during the simulation, we keep a running list of
particles that had passed through each of our $a$--$s$ bins.  The
impactor is assumed to have the same location as the target, but its
size and velocity are chosen from this running list.  However, since
the object chosen from the list was not exactly at the target's
location, we scale the magnitude of its velocity vector by
$\sqrt{a_{\rm imp}/a_{\rm targ}}$, where $a_{\rm imp}$ and $a_{\rm
  targ}$ are the instantaneous heliocentric distance of the impactor
and target respectively, and rotate it to the same longitude assuming
cylindrical symmetry.  We also assume that two particles bounce off of
one another, but that the coefficient of restitution is very small.
The end result is that we change the velocity of our target tracer to
be the mass weighted mean of its original velocity and that of the
impactor.

We have to spend a little time discussing how we decide which impactor
to choose from our running list because if this is done incorrectly it
leads to a subtle error in the results.  As described in
\citet{Levison:2007Icar:189:196}, it is imperative that our code be
able to support eccentric rings.  We found we can accomplish this if,
rather than choosing an object at random, we choose the object that
has the true anomaly that is closest to that of the target tracer.  In
this way, asymmetries can be supported by the code.  See
\citet{Levison:2007Icar:189:196} for more detail on this issue.

The growth and fragmentation of the planetesimals are also included in
LIPAD --- i.e$.$ we change $s$ in response to collisions.  When two
objects collide in nature, they produce a distribution of objects that
follows a size distribution $n(s)\,ds$ that is a function of the
physical properties of the objects involved and their relative
velocity.  Recall that LIPAD assumes that all of the planetesimals
that make up a tracer have the same size.  Thus, in order to determine
the new, post-impact, value of $s$ for the tracer involved in the
collision, we choose a random number $\psi \in \uset$ and invert the
equation
\begin{equation}
  \label{eq:SFD}
  \psi = \frac{1}{m_c}\int\limits_s^\infty m(s)\, n(s)\, ds
\end{equation}
for $s$, where $m_c \equiv \frac{4}{3}\pi\rho\left(s_{\rm
    targ}^3\!+\!s_{\rm int}^3\right)$ is the total mass of
planetesimals involved in the collision and $m(s) = {4\over 3}\pi\rho
s^3$.  {\color{MyChange} In other words, in the absence of
  fragmentation, all the planetesimals in a given target tracer would
  change to a new mass given by the sum of the old target mass plus
  projectile mass, $m_c$. With fragmentation, the new mass is chosen
  from a probability distribution that gives the fractional mass of
  collision fragments at each size. So, if $\psi = 0$ the tracer will
  be assigned the largest mass in the distribution, if $\psi = 1$ the
  smallest, and if $\psi = 0.5$, for example, then the new target mass
  is chosen so that half the total mass of fragments lies above the
  new target mass, and half below.}

It is important to note that by setting the tracer's size to this new
value of $s$ we are effectively changing {\it all} the constituent
planetesimals of the tracer to the same size.  We believe that this is
a reasonable approach in a system where a large number of collisions
are occurring.  After all, if one were to imagine that we had a large
number of collisions with exactly the same characteristics, the
procedures described above would reproduce $n(s)$.  We show in
\S{\ref{sec:tests}} that this algorithm does function well as long as
there are enough tracers in the system to adequately represent its
size distribution.

Our algorithm for determining $n(s)$ is strongly based on that used in
Eulerian statistical code developed in \citet[][hereafter
MBNL09]{Morbidelli:2009Icar:204:558}.  Following MBNL09, which itself
is based on arguments in \citet[][hereafter
BA99]{Benz:1999Icar:142:5}, we define $Q^*_D$ as the specific impact
energy (energy per unit total system mass) required to disperse 50\%
of the total mass of the interloper and target. Note that here we are
not using the tracer mass, $m_{\rm tr}$, but the combined mass of the
planetesimals $m_c$.  Defining $s_{\rm eff}$ to be the radius of an
object with this combined mass (i.e$.$ $s_{\rm eff} = \sqrt[3]{s_{\rm
    targ}^3\!+\!s_{\rm int}^3}$), MBNL09 use
\begin{equation}
  \label{Eq:qstar}
    Q^*_D = Q_0 \left(\frac{s_{\rm eff}}{1cm}\right)^\alpha +B\rho\left(\frac{s_{\rm eff}}{1cm}\right)^\beta,
\end{equation}
where $Q_0$, $B$, $\alpha$, and $\beta$ are parameters that depend on
the material properties of the objects involved.  Users are free to
set them to any values they wish in LIPAD.

Based on SPH experiments and analysis, BA99 found the $n(s)$ is best
represented by one large remnant and a continuous distribution of
fragments. MBNL09 argues that the mass of the largest remnant is:

\begin{equation}
\label{Eq:mlr}
m_{\rm LR} = \Biggm\{ \begin{array}{cc} 
    \left[-{{1}\over{2}} \left({{Q}\over{Q^*_D}}
        -1\right)+{{1}\over{2}}\right] m_c & ~~~~{\rm if~} Q<Q^*_D \\
\left[-0.35 \left({{Q}\over{Q^*_D}}
-1\right)+{{1}\over{2}}\right] m_c, & ~~~~{\rm if~} Q\geq Q^*_D \\
\end{array}
\end{equation}
where $Q$ is kinetic energy of the projectile per unit mass of the
collision.  Whenever $m_{\rm LR}$ is determined to be negative we
assume that the target is fully destroyed and we demote the tracer to
a dust particle.  

%If $m_{\rm LR} > m_{\rm targ}$ the collision is
%accretional, while it is erosional otherwise.

In all other cases, we need to determine the size-distribution of the
fragments that were formed in the collision.  To accomplish this
MBNL09 first turned to the SPH simulations in
\citet{Durda:2007Icar:186:498}. Their results typically show that the
fragments have a continuous power-law size distribution truncated at
large sizes at what MBNL09 call the largest fragment, with mass
$m_{\rm LF}$.  MBNL09 found
\begin{equation}
m_{LF}=8\times 10^{-3} ~m_c~ \left[{{Q}\over{Q^*_D}}\exp^{-\left({{Q}\over{4
Q^*_D}}\right)^2}\right]
\label{Eq:Mlf}
\end{equation}
for the mass of the largest fragment and
\begin{equation}
q=-10+7\left({{Q}\over{Q^*_D}}\right)^{0.4}\exp^{-{{Q}\over{7
Q^*_D}}}
\label{Eq:qf}
\end{equation}
for the cumulative slope of the power-law size distribution of the
fragments.

However, as MBNL09 pointed out, for all physically meaningful
collisions, the size-distribution resulting from Eqs.~\ref{Eq:Mlf} and
\ref{Eq:qf} contain infinite mass. To avoid this problem, they assume
that the fragment size distribution has a cumulative slope $q=-2.5$
\citep{Dohnanyi:1969JGR:74:2431} for $s < s_t$.  They then calculate
$s_t$ so that the integral over the resulting $n(s) m(s) ds$ is equal
to $m_c$.  We follow the same procedures here.

So, in summary, we use the total mass involved in the collision and
the impact velocity to determine $n(s)$ from Eqs.~\ref{Eq:qstar} --
\ref{Eq:qf}.  From this, we use Monte Carlo techniques to determine a
new size for the tracer.  Depending on the details of the collision,
$s$ can either increase or decrease indicating accretion or erosion.
However, the total mass of the tracer does not change as a result of
collisions.  If $s$ changes, then the number of planetesimals that the
tracer represents, $N\!=\!m_{\rm tr}/{{4\over 3}\pi\rho s^3}$, must
also change.   If $N \Longrightarrow 1$ then the
tracer is promoted to a sub-embryo.

{\color{MyChange} 

  The above algorithm is necessary but not sufficient for following
  the accumulation of objects during planet formation.  In order to
  see how it fails, we perform the following thought experiment.
  Imagine we have a system consisting of $i$ $100\,$km objects
  embedded in a population of $j$ $1\,$km objects such that $j \gg
  i$. The $100\,$km objects are represented by $k$ tracers, where $i
  \gg k$.  We also assume that the dynamics are such that growth
  occurs only by large objects accreting the small.  In this
  situation, the $100\,$km objects should increase in size while their
  number stays the same ($i$ remains constant).  In addition, as they
  grow they eat the small bodies and so $j$ decreases.  Eventually the
  big objects will run out of fuel and growth will stop.

  If we were to employ the algorithm we described above alone, we
  would get a very different behavior.  As the large objects grow, the
  number of tracers remain fixed and thus $i$ must decrease with time.
  As a result, the total mass in the large objects will remain fixed
  even though the individual objects are growing.  In addition, there
  is no transfer of mass from the small objects to the large and so
  $j$ remains constant.  So, the large objects do not run out of fuel
  and they, in principal, can grow forever.  This is obviously not
  correct.  To solve this problem, we developed the following
  algorithm for transferring mass from one size to another.

  As the system evolves, for each tracer we keep track of two
  variables that monitor how much mass it should have accreted over
  time.  The first is $f_c \equiv \prod m_f/m_i$, where the product is
  taken over all collisions that lead to growth, and $m_i$ and $m_f$
  are the initial and final mass of the constituent planetesimals at
  each collision, respectively (i.e. they are $\frac{4}{3} \pi \rho
  s^3$).  The second is an array, $\Delta m(i_s)$, that contains the
  total mass that should have been taken from each of the size bins,
  $N_{\rm s-bin}$.  These variables are reset when a tracer suffers a
  significant amount of mass loss.

  The value of $f_c$ is initially set to 1 and slowly increases with
  time.  As long as it stays less than 2, the mass deficit of the
  tracer is less than the mass of a tracer and so nothing should be
  done.  As soon as it reaches a value of 2, however, we need to
  transfer mass to this tracer (call it Tracer~$\kappa$) from smaller
  objects.  This is accomplished by giving one of the smaller tracers
  an $s$ that is similar to that of Tracer~$\kappa$.  The first step
  in this process is to choose which of the $N_{\rm s-bin}$ size bins
  to draw the mass from.  This is done by choosing a bin at random ---
  weighting the probability by $\Delta m(i_s)$.  We then choose one of
  the tracers from our running list of potential interlopers.  This
  tracer is given new size, $s = s_\kappa (1 + 10^{-3}\Gamma)$, where
  $s_\kappa$ is the size of Tracer~$\kappa$ and $\Gamma$ is randomly
  chosen from a normal distribution with a mean of zero and a
  dispersion of one.  In this way, we will now have two tracers of
  this size thereby doubling the amount of mass, and removed mass from
  the population that Tracer~$\kappa$ is growing from.  Thus, we have
  solved the problem that our thought experiment illustrated.

}

\subsubsection{Velocity Evolution of the Tracers}
\label{sssec:VEtr}

LIPAD also needs to account for the velocity evolution of the tracers.
Tracers are dynamically excited by embryos through the $N$-body
routines in SyMBA.  Indeed, this is a major advantage of LIPAD because
it will accurately handle the global redistribution of the
planetesimals by the embryos.  The tracers also affect each other's
velocity through a combination of viscous stirring, dynamical
friction, and collisional damping.  We discussed how LIPAD accounts
for collisional damping in the last section.  Here we describe its
algorithms for the two other effects.

Following standard conventions (see MBNL09, for example), we assume
that the gravitational interaction between a tracer of size $s$ and
planetesimals of smaller sizes takes the form of a drag force added to
the tracer's equation of motion.  This effect is well approximated by
the {\it dynamical friction} formalism, which, assuming a Maxwellian
velocity distribution, can be written as
\citep{Chandrasekhar1943ApJ:97:255, Binney:1987gady.book}:
\begin{equation} 
\label{Eq:df}
  \frac{d\vec{w}_m}{dt} = \frac{(m_{\rm targ}+m_{\rm
      int}) \rho_{\rm disk}}{w_m^3} \left[{\rm erf}(X) -
    \frac{2X}{\sqrt{\pi}} e^{-X^2}\right] \vec{w}_m,
\end{equation}
where $X\!\equiv\!w_m/(\sqrt{2}u$), $\vec{w}_m$ is the velocity of the
tracer with respect to the local average velocity of the small
planetesimals, $u$ is the velocity dispersion of the small
planetesimal, `erf' is the error function, and $\rho_{\rm disk}$ is
the background volume mass density of the small planetesimals.  LIPAD
calculates a separate acceleration for each of the size bins, $j_s$,
which contain smaller planetesimals.  In particular, $\rho_{\rm disk}
= m_{\rm int}~ n(j_s)$, where $n(j_s)$ is calculated with
Eq.~\ref{Eq:njs}, and $u = u_z(i_a,j_s)$.  In order to stop large
planetesimals from getting too cold, the dynamical friction
accelerations are only applied to a tracer if its relative velocity is
larger than the velocity it would have if it were in energy
equipartition with the population of small planetesimals.

Objects embedded in a disk are usually dynamically excited by larger
objects in a process known as {\it viscous stirring}.  LIPAD uses a
unique Monte Carlo algorithm to account for this mechanism between
tracers of different sizes.  In \S{\ref{sssec:rates}}, we described
how we calculate the encounter probability per timestep, $p_{\rm
  grav}$, and constructed a running list of potential interlopers in
each of our $a$--$s$ bins.  If we determine with these methods that a
scattering event occurred, we first choose an interloper at random
from the running list.  For each of the potential interlopers we
calculate $\sigma\!w$ and weigh our choice of interloper with these
values.  Recall that $\sigma$ includes gravitational focusing, and we
determine $\vec{w}$ by rotating the potential interloper to the same
location as the target assuming cylindrical symmetry.

With the interloper identified, our first step is to determine where
in the encounter plane the closest approach of the encounter occurs if
we assume no gravitational focusing.  In most cases this is
accomplished by choosing a random number, $p \in \uset$, calculating
the impact parameter,
\begin{equation}
 \label{Eq:B}
  b = r_H \sqrt{p}
\end{equation}
and a random angle $\phi \in (0,2\pi]$. We calculate the location of
the closest approach in the encounter plane, $(b_x,b_z)$, from simple
trigonometry.

However, care must be taken if one of two situations occur.  As
explained above, if the system is in the shear-dominated regime then
encounters may not be able to occur at large absolute values of $b_z$.
In particular, $\vert b_z \vert \leq h_e$ (see the discussion
associated with Eq.~\ref{Eq:sigg}).

We also have to make allowances for what MBNL09 call {\it isolated
  bodies}, which are populations of objects the are separated enough
from one another and dynamically cold enough that their orbits do not
cross.  This case imposes a lower limit on the value of $b$.  We
determine whether objects are isolated from one another in the context
of our $a$--$s$ bins, so that objects with $s \geq s_{\rm iso}$ are
isolated if
\begin{equation}
  \label{eq:iso}
   \sum_{j=j_{\rm s\_iso}}^n N_t(i_a,j) r_g(i_a, j) < \delta a,
\end{equation}
where $s_{\rm iso}$ is in bin $j_{\rm s\_iso}$, $\delta a$ is the
width of bin $i_a$, $r_g(i_a,j) \equiv c_H r_{\rm H} + 2 a
u_a(i_,j)/v_c$, $r_{\rm H}$ is the mutual Hill radius of two objects
with mass $m_p(i_a,j_s)$, $v_c$ is the circular velocity at $a$, and
$c_H$ is a parameter of the code that we set equal to $2\sqrt{3}$,
following \cite{Wetherill:1993Icar:106:190}.  If the population is
deemed to be isolated, then for each $j_s$ we define
\begin{equation}
  \label{eq:bmin}
  b_{\rm min}(i_a,j_s) = \frac{\delta a}{\sum_{j=j_s}^n N_t(i_a,j_s)} - 2 a
  u_a(i_a,j_s)/v_c.
\end{equation}
We then choose a $(b_x,b_z)$ pair so that $b_{\rm min} \leq b \leq
r_{\rm H}$ and  $\vert b_z \vert \leq h_e$.

With $(b_x,b_z)$ in hand, we can apply a kick to the velocity of the
tracer that is in response to the passage of the interloper.  Our
methods depend on the speed of the encounter.  If $w$ is faster than
the Hill velocity, $v_{\rm H} \equiv v_c
\left[(m_{targ}\!+\!m_{int})/(3M_\odot)\right]^{1/3}$, then we apply a
change in velocity calculated from the so-called {\it impulse
  velocity} approximation.  Following \cite{Spitzer:1987degc.book},
\begin{equation}
  \label{eq:vperp}
  \delta v_\perp = \frac{m_{\rm int}}{m_c} w \sin{(\gamma)}
\end{equation} 
and
\begin{equation}
  \label{eq:vpar}
  \delta v_\parallel = \frac{m_{\rm int}}{m_c} w \left[1-\cos{(\gamma)}\right],
\end{equation}
where 
\begin{equation}
     \sin{(\gamma)} = \frac{\left(\frac{2 b w^2}{m_c}\right)}{1+\left(\frac{2 b w^2}{m_c}\right)^2}.
\end{equation}
We apply $\delta v_\parallel$ along the direction of $\vec{w}$, and
$\delta v_\perp$ along the vector the connects $(b_x,b_z)$ and the
center of the target.

If $w < v_{\rm H}$ then we integrate the encounter numerically.  In
particular, we place the interloper at $(b_x,b_z)$ in a system
containing the target and the sun.  Its initial velocity is $\vec{w}$
with respect to the target.  We first move both the target and
interloper backward along their respective Kepler orbits for a time
$10 r_{\rm H} /w$, and then integrate the system forward for twice as
long.  Finally, we move the target back in time to the point of
closest approach.  Its change in velocity is calculated from its final
position and velocity.  This change is applied to the tracer at its
original position.

In addition to the viscous stirring, we must include, at least
crudely, the self-gravity in the tracers to prevent unphysical
migration of the embryos (LTD10).  We use the algorithm developed in
LTD10, which are based on a technique originally developed for the
study of disk galaxies, known as the particle-mesh (PM) method
\citep{Miller:1978ApJ:223:811}.  In what follows, we use the formalism
from \cite{Binney:1987gady.book}.  We first define a modified polar
coordinate system ${\rm u}\!\equiv\ln{\varrho}$ and $\phi$, where
$\varrho$ and $\phi$ are the normal polar coordinates, and define a
{\it reduced} potential, $V({\rm u},\phi) = e^{{\rm u}/2}
\Phi\left[\varrho({\rm u}),\phi\right]$ and a {\it reduced} surface
density $S({\rm u},\phi) = e^{{\rm u}/2}\sigma\left[\varrho({\rm
    u}),\phi\right]$ such that:
\begin{equation}
V({\rm u},\phi) = -{G\over\sqrt{2}} \int\limits_{-\infty}^{\infty}
\int\limits_{0}^{2\pi} {S({\rm u}',\phi') d\phi' \over \sqrt{\cosh{({\rm u}-{\rm u}')} -
\cos{(\phi-\phi')}}} d{\rm u}'
\end{equation}
If we break the disk into cells this becomes: 
\begin{equation}
\label{Eq:Vlm}
V_{lm} \approx
\sum\limits_{l'} \sum\limits_{m'} {\cal G}(l'-l, m'-m) {\cal
  M}_{l'm'}
\end{equation}
where ${\cal M}_{lm} = \int\int_{{\rm cell}(l.m)} S\,d{\rm u}\,d\phi$ and
${\cal G}$ is the Green's function:
\begin{equation}
  \label{eq_gr}
  {\cal G}(l'-l, m'-m) = -{G \over \sqrt{2\left(\cosh{({\rm u}_{l'}-{\rm u}_l)} -
      \cos{(\phi_{m'}-\phi_{m})}\right)}},
\end{equation}
when $l\neq l'$ and $m\neq m'$, and 
\begin{equation} {\cal G}(0,0) =
  -2G\left[{1\over{\Delta\phi}}\sinh^{-1}{\left(\Delta\phi\over\Delta
        {\rm u}\right)} + {1\over{\Delta {\rm
          u}}}\sinh^{-1}{\left(\Delta {\rm u}\over\Delta
        \phi\right)}\right],
\end{equation}
where ${\Delta {\rm u}}$ and $\Delta\phi$ are the grid spacings.

For this algorithm we found that it is best to assume that the disk is
axisymmetric, so Equation~\ref{Eq:Vlm} becomes
\begin{equation}
V_{lm} \approx \sum\limits_{l'}
\sum\limits_{m'} {\cal G}(l'-l, m'-m) {\Delta\phi\over2\pi} {\cal
  M}_{l'} = \sum\limits_{l'} {\Delta\phi\over2\pi} {\cal M}_{l'}
\sum\limits_{m'} {\cal G}(l'-l, m'-m)
\end{equation}
\begin{equation}
\label{Eq:Vl}
V_l \approx \sum\limits_{l'}
{\Delta\phi\over2\pi}
{\cal M}_{l'} \tilde{\cal G}(l'-l).
\end{equation}
Note that Equation~\ref{Eq:Vl} is one dimensional, and thus it only
supplies us with a radial force.  The tangential and vertical forces
are assumed to be zero.  We made this assumption due to the small
number of tracers in our system.  However, a simple radial force is
adequate for our purposes (LTD10).

Also, the form of Equation~\ref{Eq:Vl} allows us to use the $a$ bins
that we already constructed for the collisional algorithm.  All we
need is that relationship between ${\cal M}_{l'}$ and the total amount
of mass in ring, $M_a \equiv \sum_j M_t(l,j)$.  We find that
\begin{equation}
{\cal M}_{l'} = {2 M_a\over\left(a_{l2}^2
    - a_{l1}^2\right)} a_l^{3/2} \left[\ln(a_{l2}) -
  \ln(a_{l1})\right], 
\end{equation}
where $a_{l2}$, $a_{l1}$, and $a_l$ are the outer edge, inner edge,
and radial center of ring $l$.

So, Equation~\ref{Eq:Vl} gives us the reduced potential at the center
of ring $l$ and thus the true potential can be found ($\Phi = e^{-{\rm
    u}/2}V$).  To calculate the radial acceleration at any location,
we employ a cubic spline interpolation scheme.  Finally, the
acceleration of a particle is calculated by numerically
differentiating this interpolation.

%{\color{red}(The code currently doesn't include the potential of the gas.)}

A user also has the option of including the drag and tidal effects of
a gas disk on the particles.  For the tracers this means adding a
fictitious force that mimics aerodynamic drag.  Our basic algorithm is
described in detail in LTD10.  It includes cases where the Knudsen
number is larger than unity, which occurs when a molecule's
mean-free-path is larger than the size of the particle, using the
prescription of \citet{Adachi:1976PTP:56:1756}. This can occur for
small bodies in the outer region of the nebula.

In order to calculate the drag on particles, we need to adopt a model
for the nebula. Our model, which is based on that of
\citet{Hayashi:1985prpl.conf:1100}, has the form
\begin{equation}
\label{eq:gden}
\rho_g(\varpi,z) = \rho_{0,g} \left(\frac{\varpi}{1\,{\rm
AU}}\right)^{-\alpha} e^{-z^2/{z_s}^2(\varpi)},
\end{equation}
where $\varpi$ and $z$ are the cylindrical radius and height,
respectively, $\rho_{0,g}$ is the gas density in the plane at $1\,$AU,
and $z_s$ is the scale height of the disk at $\varpi$. The scale
height is determined by the $\varpi$-dependence of temperature $T$:
following \citet{Hayashi:1985prpl.conf:1100} we adopt $T=T_0\, (\varpi
/ {1\,{\rm AU}})^{-1/2}$ so that
\begin{equation}
z_s(\varpi) = z_{0,s} \left(\frac{\varpi}{1\,{\rm AU}}\right)^{5/4},
\end{equation}
where $z_{0,s}$ is the scale height of the disk at $1\,$AU. Their
``minimum mass '' model has $z_{0,s}=0.047$, $\alpha=2.75$, and
$\rho_{0,g}=1.4\times 10^{-9}\,{\rm gm/cm^3}$. However, the user is
free to set these parameters to whatever value they wish.  The user
also has the option to have the gas disk exponentially decay away with
a timescale of $\tau_{\rm gas}$.

Finally, in order to determine the headwind that a planetesimal will
experience, we need to determine the local circular velocity of the
gas, $v_g$, in our model. As is conventional we define
\begin{equation}
\label{eq:eta}
\eta \equiv \frac{1}{2} \left[ 1 - \left(\frac{v_g}{v_c}\right)^2\right].
\end{equation}
For our assumed temperature profile,
\begin{equation}
\eta = 6.0 \times 10^{-4} \left(\alpha+\frac{1}{2}\right) \left(\frac{\varpi}{1\,{\rm AU}}\right)^{1/2}.
\end{equation}
The instantaneous acceleration on the tracer is determined from
$\rho_g$, $v_g$, and $s$.  Following the examples of its Eulerian code
brethren, LIPAD currently does not include the acceleration due to the
gravitational potential of the gas disk.

In summary, the velocity of the tracers are affected by four
processes: dynamical friction, aerodynamic drag, self-gravity, and
viscous stirring.  The first two are included in LIPAD with the use of
analytic expressions.  Self-gravity is included through the use of the
PM method.  Viscous stirring is included via a Monte Carlo algorithm
that applies velocity kicks to the particles based on the local
characteristics of other tracers.

\subsection{Behavior of the Embryos}
\label{ssec:embryos}

\subsubsection{Velocity Evolution of the Sub-Embryos}
\label{sssec:VSEm}

The full-embryos interact with all the objects in the simulation via
the $N$-body algorithms in SyMBA.  Thus, no special routines need to
be included in LIPAD to handle their velocity evolution.  This is not
true for the sub-embryos.  Recall that we created this class of object
so that when a tracer is promoted to an embryo, it does not suddenly
find itself embedded in a disk of similar-mass objects.  Thus,
although sub-embryos interact with the full-embryos and other
sub-embryos through the $N$-body algorithms, we needed to construct
special routines to calculate the gravitational interactions between
the sub-embryos and the tracers.  We include two-types of interactions
in LIPAD: dynamical friction and planetesimal-driven migration.  For
the former we employ the analytic formalism described in
\S{\ref{sssec:VEtr}} near Eq.~\ref{Eq:df}.

Before we can explain the methods we use to include sub-embryo
planetesimal-driven migration into LIPAD, we first need to discuss
some aspects of how the process works.  A planet or planetary embryo
placed into a disk of planetesimals will migrate as a result of an
asymmetry in the way it gravitationally scatters the planetesimals.
This type of migration occurs only when conditions are right
\citep{Kirsh:2009Icar:199:197,Minton:2012Icar:sub}.  From the point of
view of designing LIPAD, the most relevant condition is that an embryo
will only migrate if the planetesimals it is interacting with are
{\color{MyChange} at least} 150 times less massive than it is.  Thus,
a sub-embryo will not migrate if it directly interacts with the
tracers and we needed to develop a way of splitting the tracers into
their component planetesimals.

In this algorithm we basically perform a series of three body
integrations, similar to those described above, that include the Sun,
sub-embryo, and a single interloper.  Each interloper is chosen from
the tracers that can get within $A r_H$ of the sub-embryo, where $A$
is a free parameter that we set to 7 in the examples below.  It has
the same semi-major axis, eccentricity, and inclination as one of the
tracers, but its phases are chosen so that it is initially in the
encounter plane at $(b_x,b_z)$.  The value of $(b_x,b_z)$ is chosen
via the procedures described near Eq.~\ref{Eq:B} above.  We move both
the target and interloper backward along their respective Kepler
orbits for a time $10\,\tau_{\rm syn}$, where $\tau_{\rm syn}$ is that
synotic period of the pair.  Then we integrate the system forward for
twice as long.  We record the change in energy of the sub-embryo's
orbit.

We determine the total number of such encounters using the same
procedures we used to determine $p_{\rm grav}$ for the tracers.  At
the end of these integrations we have the total change in energy that
the sub-embryo should experience in the next timestep, and thus
$\dot{E}$.  This energy change is smoothly added to the orbit of the
sub-embryo via a fictitious acceleration applied to its equation of
motion.  Tests of this algorithm are presented in \S{\ref{ssec:Mig_test}}

\subsubsection{Embryo-Disk Tidal Interactions}

If a growing planetary embryo is in the presence of gas, it will
migrate due to planet-disk tidal interactions
\citep{Ward:1986Icar:67:164, Korycansky:1993Icar:102:150,
  Ward:1997Icar:126:261}.  As in LDT10, we use the approach of
\citet{Papaloizou:2000MNRAS.315.823}.  This approach has the advantage
that it can handle the case where a protoplanet's eccentricity is
greater than the scale-height--to--semi-major axis ratio. They derive
timescales for semi-major axis damping, $t_a$, and for eccentricity
damping, $t_e$, for an embryo of mass $M$ at semi-major axis $a$ with
eccentricity $e$:

\begin{equation}
\label{eq:ta}
t_a = \frac{1}{c_a} \, 
\sqrt{\frac{a^3}{G M_\Sun}} 
\left(\frac{z_s}{a} \right)^{2} 
\left(\frac{\Sigma_{\rm{g}} \pi a^2}{M_\Sun}\right)^{-1} \,
\left(\frac{M}{M_\Sun}\right)^{-1} \,
\left[
\frac
{1 + (\frac{e a}{1.3 z_s })^{5}}
{1 - (\frac{e a}{1.1 z_s })^{4}}
\right]
\end{equation}

\begin{equation}
\label{eq:te}
t_e = \frac{1}{c_e} \, 
\sqrt{\frac{a^3}{G M_\Sun}} 
\left(\frac{z_s}{a}\right)^{4} 
\left(\frac{\Sigma_{\rm{g}} \pi a^2}{M_\Sun}\right)^{-1} \,
\left(\frac{M}{M_\Sun}\right)^{-1} \,
\left[1 + \frac{1}{4} \left(\frac{e a}{z_s }\right)^{3}\right]
\end{equation}
where $\Sigma_g$ (= ${\pi}^{1/2} \rho_g z_s$) is the local gas surface
density.  \citet{Papaloizou:2000MNRAS.315.823} also argue that if the
inclination damping timescale ($t_i$) is not significantly shorter
than the eccentricity damping timescale then it plays little role in
the equilibrium state; we set $t_i = t_e$ for simplicity.

From the formulas above we can find the acceleration on an object due
to tidal damping of semi-major axis and random velocity, namely
\begin{equation}
\label{eq:a}
\vec{a}_{\rm{tidal}} = -\frac{\vec{v}}{t_a} - \frac{2
(\vec{v} \! \cdot \! \vec{r})}{r^2 \, t_e}
- \frac{2 (\vec{v} \! \cdot \! \vec{k})
\vec{k}}{t_i}
\end{equation}
where $\vec{r}$, $\vec{v}$, and $\vec{a}$ are Cartesian position,
velocity, and acceleration vectors, respectively (with $r$ as the
magnitude of the radial vector) and $\vec{k}$ is the unit vector in
the vertical direction.  The user is free to set $c_a$ and $c_e$ to
whatever values they wish.

\subsubsection{Collisional Evolution of the Embryos}
\label{sssec:EmCol}

Basic SyMBA makes the simple assumption that when two objects hit one
another they perfectly merge (i.e$.$ create one object with all the
mass while conserving linear momentum).  This assumption is adopted by
all $N$-body codes, as far as we are aware.  Given the effort we put
into accurately following the evolution of the tracer
size-distribution (see \S{\ref{sssec:col}}), we feel that we need give
the user the ability to relax this assumption in LIPAD.
Unfortunately, the structure of LIPAD puts several constraints on our
ability to accomplish this.  In particular, we cannot increase the
number of particles (tracers plus embryos) in the system for fear that
the $N$ would run away and become too large to be computationally
tractable.  In addition, the statistical algorithms for the tracers
require that all the tracer particles have the same mass.  Thus, the
algorithm that we developed to handle the collisional evolution of the
embryos are necessarily cruder than those we use for the tracers, but
are based on the same principals.

When two embryos collide, we determine the value of $m_{\rm LR}$ using
Eq.~\ref{Eq:mlr}.  We define the $m_{\rm ej} \equiv m_c - m_{\rm LR}$
as the mass of the ejecta.  We distribute the total mass involved in
the collision (i.e$.$ $m_c$) based on the value of these numbers:

\newpage

\begin{description}
\item [if $m_{\rm lr} \leq m_{\rm tr}$ then]  \hfill\break 
      \vspace*{-11truemm}
      \begin{itemize}
      \item[] We declare that both objects are pulverized.  We replace
        both of them with new tracers.  One has $s = \sqrt[3]{3 m_{\rm
            lf} /(4\pi\rho)}$ and the other 
        has $s = \sqrt[3]{3 m_{\rm
            lr} /(4\pi\rho)}$, where $m_{\rm lr}$ and $m_{\rm lf}$ are
        defined in Eqs.~\ref{Eq:mlr} and \ref{Eq:Mlf}, respectively.
        Note that this is the one case where the mass of the system is
        not conserved.  We believe that this is a reasonable tack to
        take given that the pair is blown apart by the collision.
      \end{itemize}
      \vspace*{-4truemm}

\item [else if $m_{\rm ej} < m_{\rm tr}/2$ then] \hfill\break 
      \vspace*{-11truemm}
      \begin{itemize}
      \item[] In this case there is not enough material in the ejecta
        to create a new tracer.  Therefore, we perfectly merge the two
        objects.
      \end{itemize}
      \vspace*{-4truemm}

\item [else, if $m_{\rm lr} \leq m_c/2$ then] \hfill\break 
      \vspace*{-11truemm} 
      \begin{itemize}
      \item[] We are faced with the awkward situation that most of the
        mass is in fragments.  Given our constraint that we do not
        want to increase $N$, we simply set the mass of both embryos
        to $m_c/2$.
      \end{itemize}
      \vspace*{-4truemm}

\item [else, if $m_{\rm ej} > m_{\rm tr}$ then] \hfill\break 
      \vspace*{-11truemm} 
      \begin{itemize}
      \item[] The one embryo is given the mass of $m_{\rm
          lr}$ and the other $m_{\rm ej}$.
      \end{itemize}
      \vspace*{-4truemm}

\item [else] \hfill\break 
      \vspace*{-11truemm}
      \begin{itemize}
      \item[] The one embryo is given the mass of $m_c - m_{\rm tr}$
        and the other becomes a tracer with $s = \sqrt[3]{3 m_{\rm lf}
          /(4\pi\rho)}$.
      \end{itemize}
      \vspace*{-4truemm}
\item [endif] 
\end{description}

\noindent Clearly, in the above algorithm we make a few serious
simplifications in order to satisfy LIPAD's restrictions.  However,
our choices are reasonable and are clearly preferable to the perfect
accretion that other $N$-body codes employ.  It is important to note
that in all but one of the cases above, mass is conserved by the
algorithm.  The first case is the only exception, and it only occurs
when the impact velocity is significantly larger than the escape
velocity of the pair.  Given that in most situations, the impact
velocity is set by the embryos themselves, we expect this case to
occur rarely.  Indeed, it never occurred in the test cases given in
\S{3}.

\subsubsection{Embryo Atmospheres and the Accretion of Planetesimals}

In LTD10 we presented a review of the problems the community faces
when trying to build the cores of the giant planets before the gas
nebula dissipates.  In the last decade or so, there has been a
concerted effort by the planet formation community to overcome these
problems.  This has led to the development of some additional
mechanisms intended to enhance the growth rates of planetary embryos.
One particularly promising method was developed by
\citet{Inaba:2003A&A:410:711}.  They show that the effective capture
cross-section of an embryo is significantly increased by the presence
of an extended atmosphere that is accreted from the surrounding
nebula. In LIPAD, we supply the user with the option to mimic this
effect.  In particular, we employ the formalism developed by
\citet{Chambers:2006ApJ:652L:133}, who showed that, assuming the
relative velocity of the particles is small compared to the escape
velocity of the embryo and that the scale height of the atmosphere is
set by the energy input due to accreting planetesimals, the effective
accretion radius ($R_{\rm C}$) of an embryo is
\begin{equation}
\label{eq:rc}
R_{\rm C}^4 = 0.0790\, \frac{\mu^4\, c\,
R^5\, r_H}{\kappa\,
s\, {\dot m}_R} \left(\frac{M}{M_\odot}\right)^2,
\end{equation}
where $R$ and $s$ are the radius of the embryo and planetesimal,
respectively, $M$ is the embryo's mass, $\mu$ is the mean molecular
weight of the atmospheric gas, $\kappa$ is its opacity, and $c$ is the
speed of light. The parameter ${\dot m}_R$ is the accretion rate that
the embryo would have had if there was no atmosphere. We calculate
this value for each embryo in real time during our simulation by
monitoring the number of tracer particles that pass through the
embryo's Hills sphere, and extrapolating to its surface. During our
simulations, we do not allow $R_{\rm C}$ to exceed $0.5\,r_H$.  If an
incoming tracer comes within a radius $R_{\rm C}$ of the embryo, it is
assumed to have a collision.

One issue that we need to address concerning the embryo atmospheres is
how they evolve as the gas disk decays and embryos collide with one
another.  First note that Eq.~\ref{eq:rc} is independent of the
surface density of the gas disk.  This is due to the fact that the
most important regulator for the gas accretion rate onto an embryo is
the heating and cooling of the atmosphere, itself.  The atmosphere is
gravitationally bound to the planet and pressure supported.  It needs
to cool and collapse before more gas can be added.  Thus, given that
the atmosphere is already bound to the embryo, we expect the
atmospheres to survive once the gas disk dissipates --- at least in
the absence of giant impacts.  In that regard,
\citet{Genda:2003Icar:164:149} found that planets only lose $\sim\!20$
--- $\sim\!30\%$ of their atmospheres during such collisions.

The fact that only a couple of tens of percent of atmospheres are lost
during a collision and it is not clear how this loss affects $R_{\rm
  C}$ (which is mainly determined by thermal evolution), we decided to
take the simplistic approach of allowing the atmospheres to survive in
our simulations for all time.  We adopt Eq.~\ref{eq:rc} even after the
gas disk dissipates and after giant collisions.  We feel that this
assumption is reasonable given that the user has the option in LIPAD
of ignoring the atmospheres entirely.  Thus, we believe it best for
models that include atmospheres to represent the end-member of
possible simulations that maximize the effects of those atmospheres.
This is the approach we have taken.

\medskip\medskip\medskip

This concludes our description of LIPAD.  In the previous sections we
described the three major classes of objects.  Each object in the code
gravitationally and collisionally interacts with its neighbors, which
leads to changes in the dynamical state of the system as well as the
sizes of the objects involved.  This is the first code of which we are
aware that can accurately follow the evolution of a system initially
containing only small planetesimals, as objects grow and evolve, until
they reach fully formed planets.  In the next section we present tests
of LIPAD.

\section{Tests of LIPAD}
\label{sec:tests}

LIPAD has been carefully verified and tested. In this section we
present some of these tests.  Tests were chosen that not only verify
that the code is behaving properly, but also illustrate how the code
works.

\subsection{Collisional Damping}

The first test we present is one designed to test a combination of our
particle-in-a-box algorithms for determining the collision rates
between tracers (\S{\ref{sssec:rates}}) and the collisional damping
routines (\S{\ref{sssec:col}}).  Following MBNL09 (see their
Figure~12), we calculate the evolution of a system of $10\,$km
planetesimals spread from 4 to $6\,$AU. The system contains
$32\,M_\oplus$ of material. In LIPAD, we represented this population
with 1660 tracer particles, and so each tracer initially represents
9.6 million planetesimals. We only included collisional damping in
this calculation --- particles did not fragment or accrete.

\begin{figure}[h!]
\includegraphics[scale=0.5]{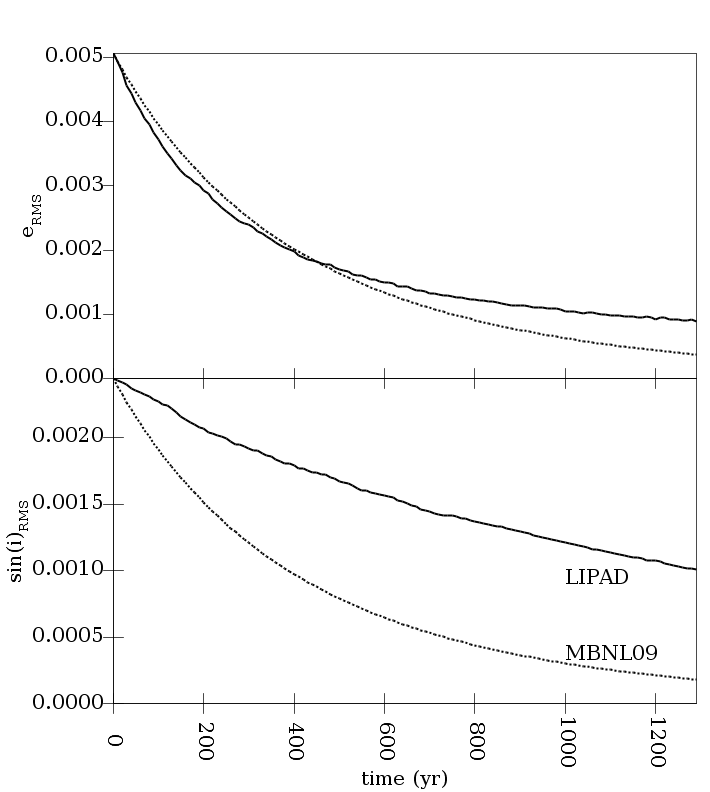}
\caption{\footnotesize \label{fig:cdamp} The time evolution of the RMS
  eccentricity and sine of the inclination of a system containing $1.6
  \times 10^{10}$ or $32\,M_\oplus$ of $10\,$km objects. These objects
  were spread from 4 to $6\,$AU from the Sun.  The solid curve
  represent the results from a LIPAD simulation consisting of 1660
  tracers.  The dotted curve shows the results from MBNL09's
  statistical grid code.}
\end{figure}

Figure~\ref{fig:cdamp} shows a comparison of our LIPAD results (solid
curves) to those from MBNL09's Eulerian code (dotted curves).  The
eccentricity evolution is quite similar in the two codes, especially
at the beginning of the simulation when the two systems are in the
same dynamical state. However, the inclination damping is much slower
in LIPAD.  In particular, $e_{\rm RMS}/i_{\rm RMS}$ drops from 2 to
roughly 1.2 in 1000 years.  MBNL09's code assumes that this ratio is
fixed at 2. The issue now becomes determining which of these codes are
correct.

Unfortunately, as far as we are aware, there is no work that studied
the behavior of a collisionally damped system where the coefficient of
restitution is zero (which is the case here). If it were non-zero then
physical scattering events will redirect velocities from one direction
to another because the collisions are typically off-center.  This
effectively couples $e$ and $i$ and thus $e_{\rm RMS}/i_{\rm RMS} =
2$.  In the case where the coefficient of restitution is zero,
however, $e$ and $i$ are decoupled and so it is not clear how they
will behave.

In addition, as far as we are aware, there is no code that can perform
this calculation without making serious compromises.  Fortunately, we
have a code that can, at least, mimic collisions with a coefficient of
restitution of zero --- the perfect merger routines in SyMBA lead to
the same velocity evolution.  The issue here is that SyMBA cannot
possibly handle the roughly $10^{10}$ planetesimals implied by this
calculation.  We can simplify the simulation by significantly
decreasing the number of particles, while increasing their physical
cross-section in order to decrease the impact probability and thus the
computation time.  In this way, although we do not recreate the
timescales of the true system, we can determine how $e_{\rm
  RMS}/i_{\rm RMS}$ changes during the simulation.

\begin{figure}[h!]
\includegraphics[scale=0.5]{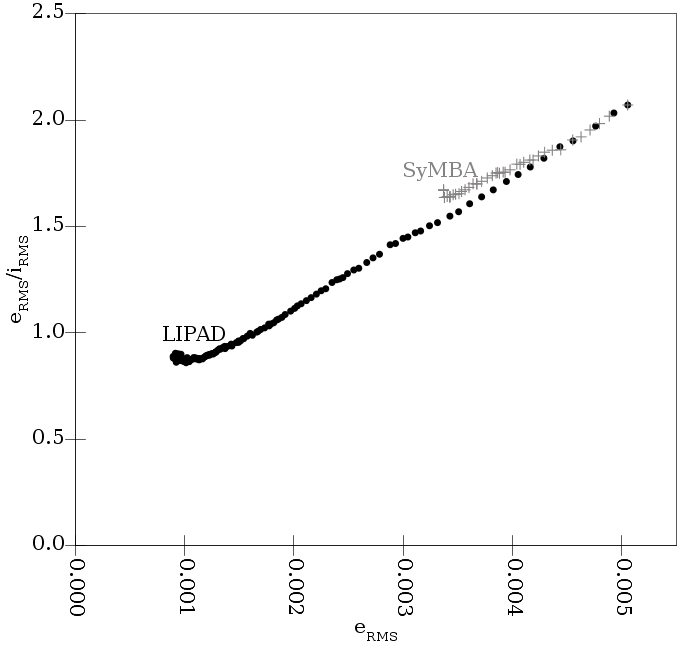}
\caption{\footnotesize \label{fig:cd_etoi} A comparison between the
  evolution of $e_{\rm RMS}/i_{\rm RMS}$ in the LIPAD run shown in
  Fig.~\ref{fig:cdamp} (black circles) and a SyMBA simulation
  initially containing the same number of objects (gray $+$'s) during
  a test of collisional damping.  The size of the particles in the
  SyMBA runs are inflated in order to suppress viscous stirring. Time
  runs from right to left in the figure.  There is excellent agreement
  between the two simulations suggesting that the LIPAD results shown
  in Fig.~\ref{fig:cdamp} are more accurate than those of MBNL09's
  code, which assume that $e_{\rm RMS}/i_{\rm RMS} = 2$.}
\end{figure}

In particular, we started with the 1660 particles in the original
LIPAD run, set their physical radius to $0.01\,$AU, and set their mass
to zero so to suppress viscous stirring.  In Figure~\ref{fig:cd_etoi}
we plot $e_{\rm RMS}/i_{\rm RMS}$ as a function of $e_{\rm RMS}$ for
our new SyMBA run (gray `$+$') and our full LIPAD runs (black dots).
By plotting $e_{\rm RMS}$ on the abscissa we remove any issues caused
by different collision timescales.  Since eccentricity monotonically
decreases during the simulation, time runs from upper right to lower
left in the figure.  Recall that $e_{\rm RMS}/i_{\rm RMS}$ remains
equal to 2 in MBNL09's code.  As can be seen in the figure, the drop
in $e_{\rm RMS}/i_{\rm RMS}$ seen in the LIPAD run is also seen in
SyMBA.  Thus, we conclude that this drop is real and that LIPAD is
correctly modeling collision damping.  Indeed, it is more accurate
than the analytic equations used by the Eulerian codes.

\subsection{Collisional Fragmentation and Accretion}

\begin{figure}[h!]
  \includegraphics[scale=0.5]{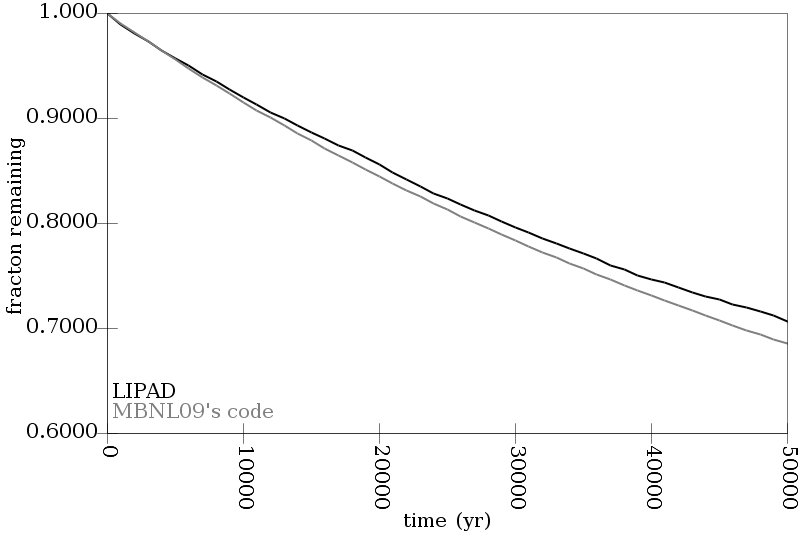}
  \caption{\footnotesize \label{fig:grind} The temporal evolution of
    the mass of a fragmenting system initially containing
    $1.3\,M_\oplus$ or $\sim\!24$ million 30-km objects spread from
    2.25 and 2.75 AU. The collisions are energetic enough that objects
    involved in a collision are converted directly to dust.  The LIPAD
    runs, which consists of 10,000 tracers, is shown in black, while a
    runs using MBNL09's Eulerian code is in gray.}
\end{figure}

As a first test of the collisional fragmentation/accretion evolution
algorithms we study a system that should quickly grind down.  We start
with a $1.3\,M_\oplus$ disk of material between 2.25 and $2.75\,$AU
made up of $30\,$km objects. We set the initial eccentricities of this
population to 0.2, which is large enough that when two objects hit,
they are pulverized in the sense that they go directly to dust. In
Figure~\ref{fig:grind} we compare the results of LIPAD to those from
MBNL09's code.  We find excellent agreement.

We follow the above simple test with one that includes both accretion
and fragmentation of the tracers (\S{\ref{sssec:col}}).  It was
inspired by MBNL09's main result that asteroids were born big. We
start with a population of $s\!=\!50\,$km objects spread from 2 to
$3\,$AU with a total mass of $1.6\,M_\oplus$. We made a couple of
modifications to the original MBNL09 calculations that are designed to
better exercise LIPAD.  First, we did not include velocity evolution
since we are only interested in studying the
size-distribution/collision-rate part of the code. In addition,
$e_{\rm RMS}$ was set to the particles' Hill eccentricity in the
original calculation. Here, we set the RMS eccentricities and
inclinations to much larger values (0.01 and 0.005, respectively) so
that we would produce a significant amount of collisional grinding ---
we required a test that included both growth and fragmentation. We
represented this system with 20,000 tracers.

\begin{figure}[h!]
  \includegraphics[scale=0.4]{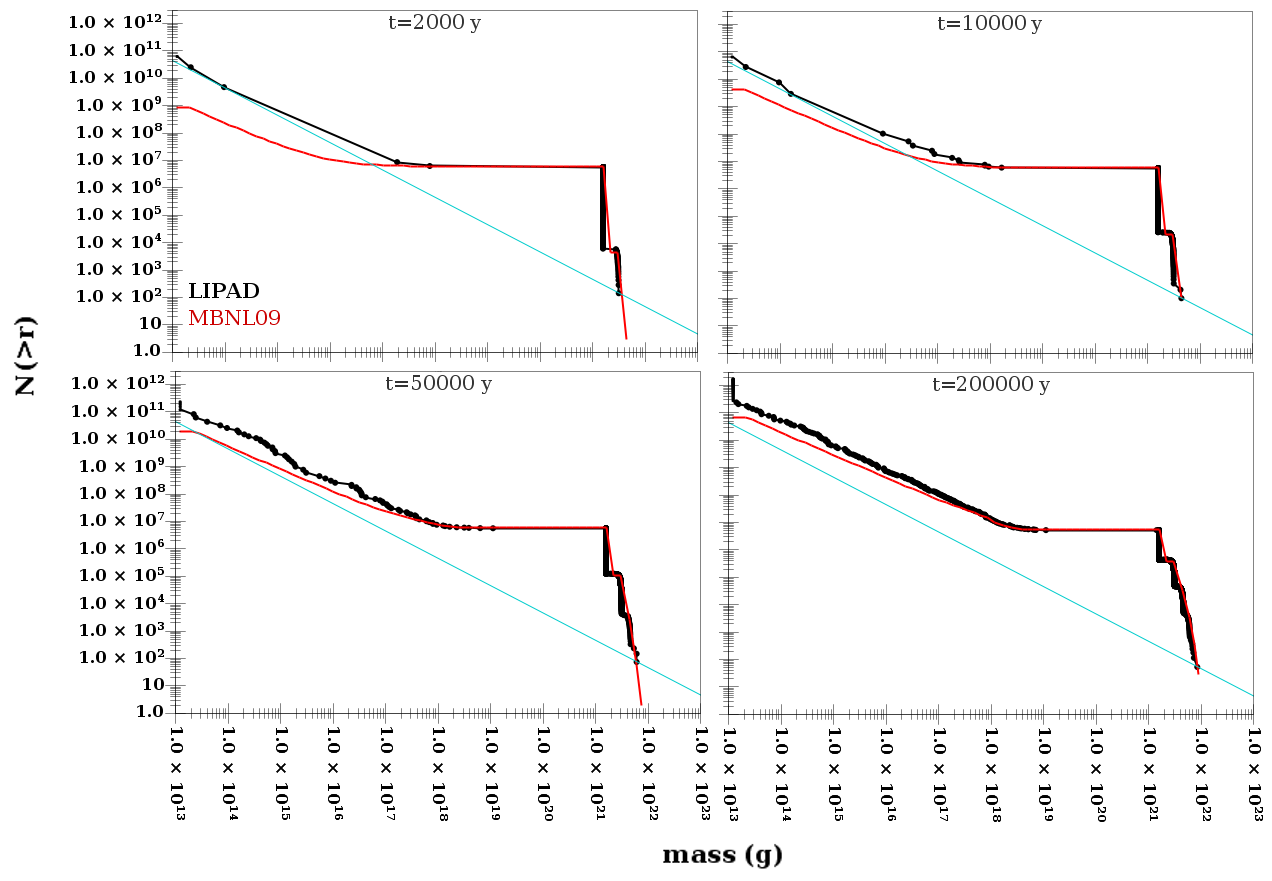}
  \caption{\footnotesize \label{fig:cevol_sfd} Four snapshots of the
    evolution of the cumulative mass distribution of a system
    initially consisting of a population of 6.4 million $s\!=\!50\,$km
    objects spread from 2 to $3\,$AU.  A $50\,$km object has a mass of
    $1.5 \times 10^{21}\,$g.  The RMS eccentricities and inclinations
    were 0.01 and 0.005, respectively, and there was no velocity
    evolution in the calculations.  The black and red curves indicate
    the results from LIPAD and MBNL09's Eulerian code, respectively.
    The cyan line shows the number of particles that a single tracer
    represents as a function of the mass of its constituent
    planetesimals.}
\end{figure}

Figure~\ref{fig:cevol_sfd} shows four snapshots of the cumulative mass
distribution according to LIPAD (black) and MBNL09's code (red).
First, we note that at all four times the size distribution produced
by the two codes for the objects that are growing (i.e$.$ for masses
greater than $10^{21}\,$g) are in excellent agreement with one
another.  However, at early times, LIPAD does not correctly reproduce
the collisional tail of the distribution.  This shows what, in our
view, is the most significant limitation of LIPAD --- the relatively
grainy resolution of the size-distribution. Recall that as the system
evolves, a tracer's mass remains constant and so, as the mass of the
planetesimals that it represents (i.e$.$ $m_p =
\frac{4}{3}\,\pi\,\rho\,s^3$) changes, the number of particles it
represents also changes. Thus, at each value of planetesimal mass,
there is a minimum number of planetesimals that a tracer can
represent, i.e$.$ $m_{\rm tr}/m_p$. This is the cyan curve in the
plot. Note that the black curve never falls below this curve.

At early times, as the collisional tail starts to develop, LIPAD
cannot represent it very well because of this resolution limitation
(note that the tail of the size-distribution from MBNL09's code lies
below the cyan line). As a result, LIPAD's tail sits far from that
produced by MBNL09's code. Note that, although the size-distribution
has a different shape, the total amount of mass in the collisional
tail is the same for the two codes during this time. LIPAD soon
recovers, however, and once the mass in the tail becomes significant,
the size-distributions match well.

We want to emphasize that, despite this limitation, LIPAD correct
reproduces the growing embryos.  At no time does the shape of the
collisional tail affect the growth of the planets.  Thus, we conclude
that, although we would not use LIPAD to study the details of the
evolution of the shape of a size-distribution during a collisional
cascade, it is very capable of following the accretion of any planets
in the system, as well as any mass loss due to collisional grinding.

   \begin{figure}[h!]
       \includegraphics[scale=0.5]{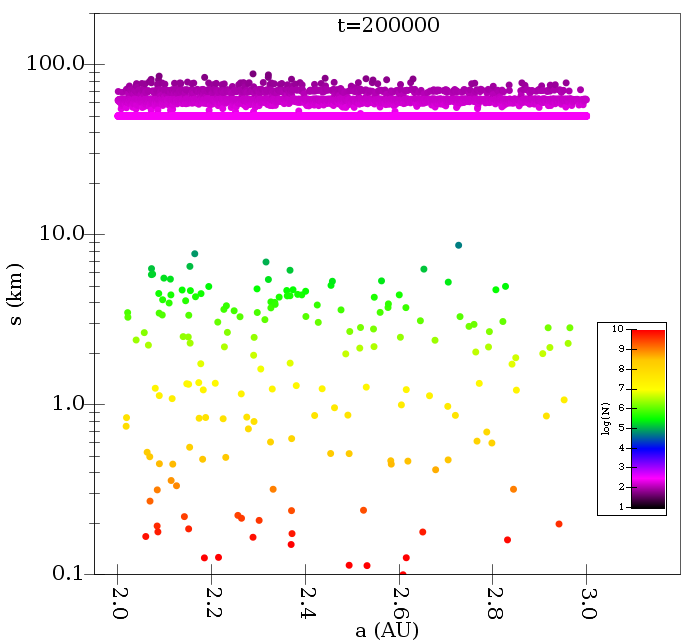}
       \caption{\footnotesize \label{fig:cevol} The last frame of an
         animation showing the temporal evolution of the tracers in
         the simulation presented in Fig.~\ref{fig:cevol_sfd}.  In
         particular, we plot the size of a tracer's constituent
         planetesimals, $s$, as a function of semi-major axis, $a$.
         Color shows the number of planetesimals that each tracer
         represents.  The full animation can be found at {\tt
           http://www.boulder.swri.edu/$\sim$hal/LIPAD.html}.}
   \end{figure}

In Figure~\ref{fig:cevol} we show what is actually happening in the
code.  Each point represents an individual tracer.  Location on the
dot shows the semi-major axis ($a$) and planetesimal size ($s$).  Its
color shows the number of planetesimals that that tracer represents.
Recall that since $m_{\rm tr}$ is fixed, as $s$ decreases this number
increases.

{\color{MyChange}
\subsection{Runaway Growth}
\label{ssec:RG_test}

As we described in \S{\ref{sec:intro}}, the process of runaway growth
has been shown to be important in the formation of the planets
\citep{Wetherill:1989Icar:77:330, Greenberg:1978Icar:35:1}.  Thus, it
is essential to determine whether LIPAD can reproduce this process.
Fortunately, there are analytic solutions to the coagulation equation
that include it, which we can use of as a test for LIPAD.  In
particular we will employ a solution by \citet[][hereafter
W90]{Wetherill:1990Icar:88:336} that includes a severe form of runaway
growth.

The coagulation equation follows the evolution of the size- or
mass-distribution of a population of objects under the assumption that
when two objects hit they merge (i.e$.$ there is no fragmentation).
W90 found an analytic solution to the discrete form of coagulation
equation where he assumes that there is a single runaway object.
Under the assumption that at any time the system can be represented by
a population of objects that are members of a continuous
size-distribution (represented by a series of mass bins with mass
$m_k$ that contain $n_k$ objects) and a runaway with mass $m_R$, then
this equation has the form
\begin{equation}
 \label{Eq:w90a}
\frac{dn_k}{dt} = \frac{1}{2}\sum\limits_{i+j=k}A_{i,j}n_in_j - 
n_k\sum\limits_{i=1}^\infty A_{i,k}n_i - A_{R,k}n_k, 
\end{equation}
where $A_{i,j}$ is the probability that an object in bin $i$ will
impact an object in bin $j$ per unit time and is, in general, a
non-linear combinations of the physical parameters of the system such
as masses, velocities, volumes, and bulk densities.  The first term in
the equation represents objects that are undergoing mergers thereby
entering bin $k$.  The second term represents those objects initially
in bin $k$ that are undergoing collisions and thus leave the bin.  The
third term are those objects initially in bin $k$ that collide with
the runaway.

W90 found a solution to the above equation for systems that initially
consisted of population of $n_0$ objects of the same mass, $m_0$, and
where $A_{ij} \equiv \gamma \frac{m_i}{m_0}\frac{m_j}{m_0}$
($m_i=im_0$ is the mass in bin $i$ and $\gamma$ is a constant).  It is
important to note that this is not a physically realistic situation
because we expect that in the absence of gravitational focusing the
$A$'s should be proportional to the physical cross-section, which is,
in turn, $\propto m^{2/3}$.  Indeed, it is an extreme version of
runaway growth because the largest object will grow much faster then
its neighbors in this case than in a more physically realistic
situation.  As such, it makes for an excellent test of LIPAD.
Following \cite{Trubnikov:1971DAN:196:1316}, W90 found that for the
continuous distribution
\begin{equation}
 \label{Eq:w90sc}
n_{k}(\eta) = \frac{n_0\left(2k\right)^{k-1}}{k!k} 
\left(\frac{\eta}{2}\right)^{k-1} e^{-k\eta},
\end{equation}
where $\eta$ is a normalized time equal to $\gamma n_0 t$, and the
mass of the runaway is
\begin{equation}
 \label{Eq:w90sr}
 m_R(\eta) = m_{tot} - \sum\limits_{k=1}^{\infty} n_k(\eta) m_k,.
\end{equation}
where $m_{tot} \equiv n_0m_0$ is the total mass of the system.  The value of
$m_R$ is zero for $\eta<1$, but increases quickly when $\eta>1$.

While these equations have a nice compact form, it turns out that they
are very difficult to evaluate numerically.  Note that for the problem
below, $k$ needs to be as large as $70,000,000$ when $\eta=1$ in order
to calculate all values of $n_k$ that are larger then $1$, and
significantly larger in order for the second term in
Eq.~\ref{Eq:w90sr} to converge.  After much effort we have been able
to develop techniques to calculate Eq.~\ref{Eq:w90sc} when $n_k > 1$
for all $\eta$'s.  Unfortunately, we have failed to find solutions to
Eq.~\ref{Eq:w90sr} near $\eta=1$ when $m_R \ll m_{tot}$ because of a
combination of the convergence issue and that fact that the equation
requires that we take the difference of two nearly equal numbers.  In
LIPAD, we represent the system with a significant number of tracers
and thus the tracer representing the runaway is promoted to an embryo
when $m_R = m_{\rm tr} \ll m_{tot}$, which is where we cannot solve
Eq.~\ref{Eq:w90sr}.  We must stop the LIPAD simulation at this point
in order to preserve our non-physical $A$'s.  As a result, when
comparing the results of LIPAD simulations to these W90 solutions, we
cannot test whether LIPAD's runaway is growing at an appropriate rate.
Fortunately, there are other quantitative comparisons to be made.

\begin{figure}[h!]
   \includegraphics[scale=0.4]{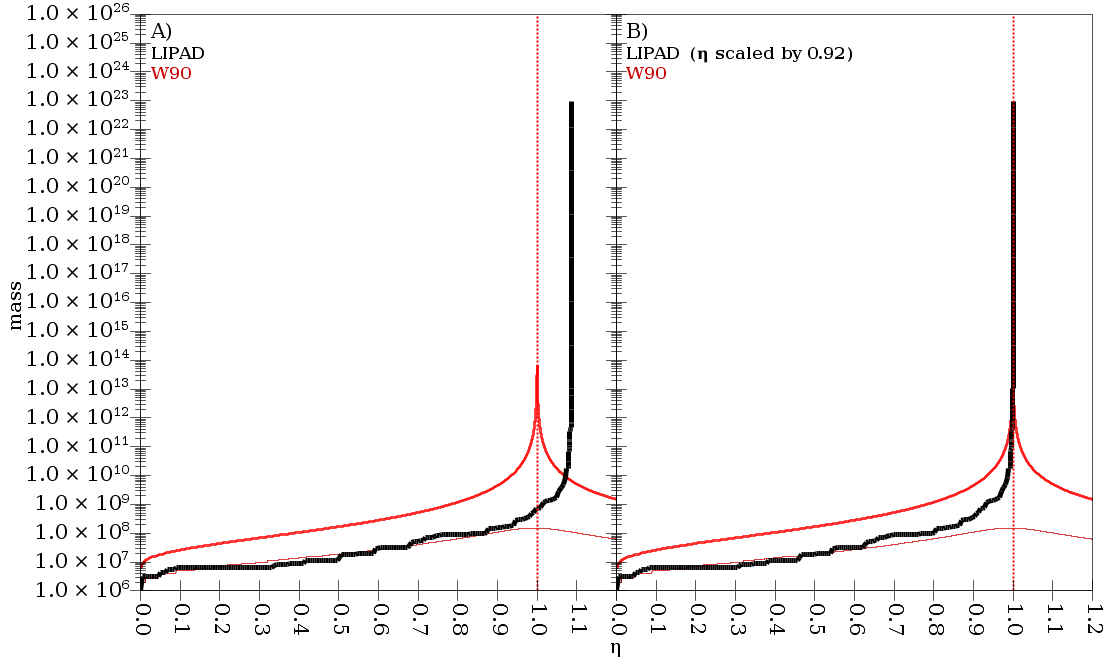}
   \caption{\footnotesize \label{fig:RW_Mmax} The temporal evolution
     of our test case based on W90 for a system initially containing
     $10^{20}$ objects of $10^6\,$g.  The black curve in (A) shows the
     mass of the largest object in our LIPAD simulation as a function
     of normalized time, $\eta$.  In (B) this curve was scaled by
     multiplying $\eta$ by 0.92.  The red curves show the analytic
     solutions for this problem.  In particular, the heavy weight
     shows the largest object in the continuous distribution according
     to Eq~\ref{Eq:w90sc}.  The vertical dotted line shows $\eta=1$,
     which is the time when runaway growth should start.  Finally, the
     thin curve shows the largest object for which $n_km_k > m_{\rm
       tr}$, which is the largest object in the continuous population
     that LIPAD can resolve (see Figure~\ref{fig:RW_snap}).  The red
     curves are the same in (A) and (B).}
\end{figure}

W90 studied a system that initially contained a population of
$10^{20}$ objects each with a mass of $10^6\,$g ($40\,$cm radius). We
adopt this test.  We present the results of the evolution of this
system in two ways.  The red heavy weight curve in
Figure~\ref{fig:RW_Mmax} shows the evolution of the mass of the
largest planetesimal in the continuous distribution, $m_l$, as a
function of $\eta$.  In particular, we define the largest mass in the
continuous distribution as the largest $m_k$ for which $n_k>1$
according to Eq~\ref{Eq:w90sc}.  The dotted lines show where $\eta=1$.
The red curves in Figure~\ref{fig:RW_snap} show snapshots of the
cumulative mass-distribution of the continuous population.

The system evolves in the following way according to W90. Initially
all objects were $10^6\,$g. When $\eta \lesssim 1$, the objects remain
part of the continuous size-distribution as they grow.  At $\eta=1$,
$m_R$ becomes non-zero and the runaway phase commences.  This occurs
when the largest object in the continuous distribution is $6.9 \times
10^{13}\,$g.  This, presumably, is the initial mass of the runaway.
The runaway grows approximately exponentially after this because
${\dot m_r} \propto {m_r}$.  Eventually, its growth slows because it
starts running out of fuel.  The value of $m_l$ (thick curve in
Figure~\ref{fig:RW_Mmax}) decreases during this time because $A_{Rj}
\propto m_j$ and thus the runaway preferentially accretes the larger
objects.

   \begin{figure}[h!]
       \includegraphics[scale=0.45]{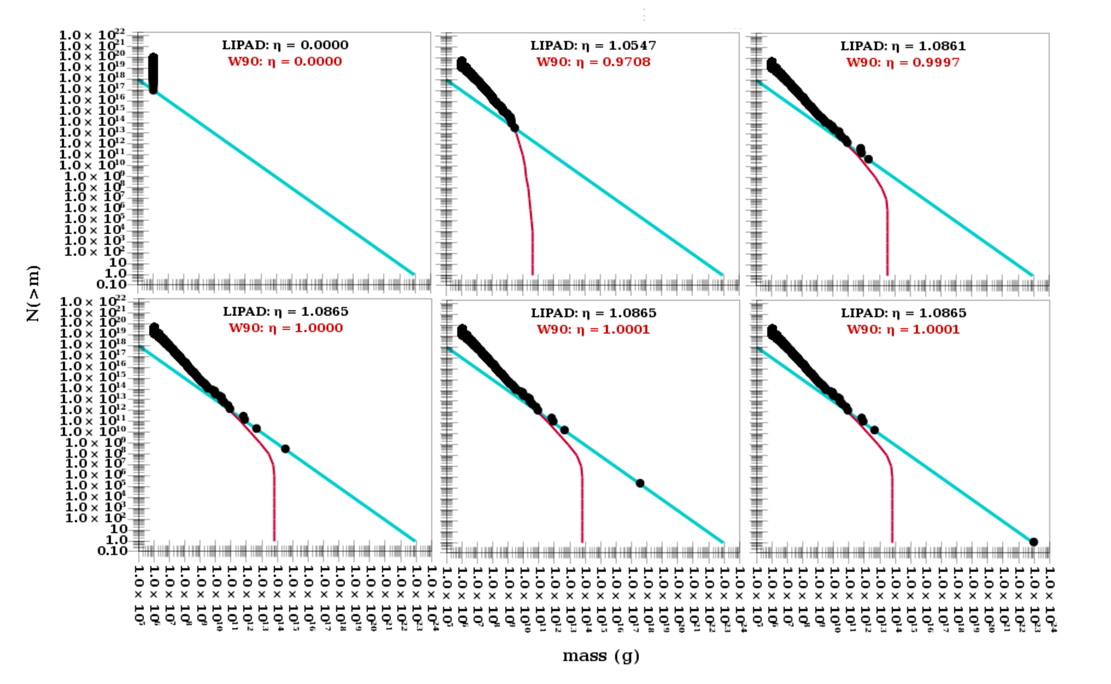}
       \caption{\footnotesize \label{fig:RW_snap} Six snapshots of the
         evolution of the cumulative mass-distribution of a system
         initially consisting of $10^{20}$ objects of $10^6\,$g.  This
         system was based on the runaway growth test in W90.  Note
         that time is highly compressed at the end of the simulation.
         The back and red curves show the results from LIPAD and an
         analytic solution, respectively.  In particular, the red
         curve shows the evolution of the continuous size-distribution
         (Eq~\ref{Eq:w90sr}). The cyan line shows the number of
         particles that a single tracer represents as a function of
         the mass of its constituent planetesimals.  As we describe in
         the text, there is a 9\% difference in the growth rates
         between the LIPAD and analytic solutions.  In order to
         compare the two size-distribution, we have remove this
         difference in this figure.  Note that time is highly expanded
         on the lower panels.  The full animation of this can be found
         at {\tt http://www.boulder.swri.edu/$\sim$hal/LIPAD.html}.}
   \end{figure}

We represented this system with 1000 tracers in LIPAD, so $m_{\rm tr}
= 10^{23}\,$g, and set $N_{\rm s-bin} = 30$.  The code itself was
customized in three ways for this calculation.  Firstly, we disabled
all the dynamical subroutines in the code.  Then we modified $p_{\rm
  col}$ to allow for the fact that $A_{ij} = \gamma
\frac{m_i}{m_0}\frac{m_j}{m_0}$.  Finally, since we force the impact
rate onto the largest object to scale as its mass, which varies from
$10^6\,$g to $> 10^{23}\,$g, we included a variable timestep in the
code.  It is important to note that the global timestep for dynamical
part of LIPAD cannot be changed because it would break the symplectic
character of the underlining SyMBA routines.  Thus, we cannot vary the
timestep of the collision code as long as we include collisional
damping in the dynamics.  We do not expect this timestep issue to
arise in a realistic situation because here the $A$'s are extreme in
that they scale as $m_im_j$, and W90 starts with very small particles
($40\,$cm), which significant increases the dynamic range of the
problem.  LIPAD does print diagnostics that will allow the user to
determine whether the timestep is becoming an issue during the
calculation.

The purpose of this test is to determine whether LIPAD can handle
runaway growth.  The black curve in Figure~\ref{fig:RW_Mmax} shows the
mass of the largest object in the LIPAD calculation as a function of
$\eta$.  As can be seen, this curve becomes almost vertical at
$\eta=1.09$.  In addition, Figure~\ref{fig:RW_Mmax} shows that at the
end of the LIPAD simulation (which occurs when the largest tracer is
promoted) we have a situation analogous to W90's prediction --- a
continuous population of objects with masses less than roughly $7
\times 10^{13}\,$g and a single, detached object that is significantly
larger ($10^{23}\,$g in this case).  Therefore, we can conclude that
LIPAD does indeed allow runaway growth to occur.

Having said this, there is a caveat we should discuss --- the timing
of the runaway.  In our LIPAD calculation, the runaway occurs at
$\eta\!\sim\!1.09$ rather than at $\eta=1$ as it should (see
Figure~\ref{fig:RW_Mmax}). Although this error is small enough (only
9\%) that we do not think it would significantly affect the results of
any real calculations, it deserves an explanation --- particularly
given that it provides insight into LIPAD.  We believe that this
offset it a result of the way in which LIPAD resolves the
mass-distribution.

The black and red curves in Figure~\ref{fig:RW_snap} show the temporal
evolution of the mass-distribution of the continuous population
according to the analytic theory (Eq.~\ref{Eq:w90sc}) and LIPAD,
respectively.  As described previously, because each tracer represents
a fixed mass, at each value of planetesimal mass, there is a minimum
number of real objects that a tracer can represent. This is the cyan
line in the plot.  Although the black curves follow the red ones
remarkably well, the former are truncated at the point where the red
curves cross the cyan line as a result of this resolution limitation.
This effect can also be seen in Figure~\ref{fig:RW_Mmax}, where the
thick red curve represents the largest mass in the continuous
distribution according to Eq.~\ref{Eq:w90sc}, but the thin curve
represents the mass where $n_km_k = m_{\rm tr}$.  Before the runaway,
the LIPAD results follow the thin curve in this figure.

As a result, at any time before runaway, the largest object in the
LIPAD simulation is smaller than the analytic theory would predict.
As discussed above, while W90's theory predicts that runaway should
occur at $\eta = 1$, it also states that the mass of the largest
planetesimal at that time is $7 \times 10^{13}\,$g.  In the LIPAD
simulation, runaway starts at a later time, but when the largest
object has roughly the same mass (see the lower left panel of
Figure~\ref{fig:RW_snap}).  Indeed, it makes sense that the onset of
runaway should be determined by mass and not time.  So, it seems
reasonable to conclude that the delay in runaway seen in LIPAD is due
to tracer resolution --- it simply takes a little longer to build an
object that can runaway.  We tested this idea further by performing a
simulation with 5000 tracers and found that runaway starts at $\eta =
1.05$ rather than $1.09$ --- a result that is consistent with this
idea.  In both our simulations, runaway starts with the largest
planetesimal is roughly $7 \times 10^{13}\,$g.  We consider the fact
that this is consistent with the predictions of W90 a success of our
code and remind the reader that the delay is less than 10\%.

}

\subsection{Viscous Stirring}
\label{ssec:VS_test}

We performed two tests of the viscous stirring routines in LIPAD
(\S{\ref{sssec:VEtr}}).  The first studies a system in the
dispersion-dominated regime and is a repeat of the test performed by
MBNL09 (see their Figure~5). In particular, we study the behavior of
$0.26\,M_\oplus$ of material in an annulus from 0.94 to
$1.06\,$AU. This annulus was populated by objects with $s = 540\,$km.
The RMS eccentricity and inclination of the disk was initially set to
$3 \times 10^{-5}$.  We performed 3 simulations using: 1) LIPAD, 2)
MBNL09's code, and 3) SyMBA for a direct $N$-body simulation. The
direct $N$-body simulation required 800 objects, while we represented
the disk with 200 tracers in LIPAD.  Fragmentation, accretion, and
collisional damping are turned off in this simulation; the only
physical effect included is viscous stirring.

\begin{figure}[h!]
   \includegraphics[scale=0.5]{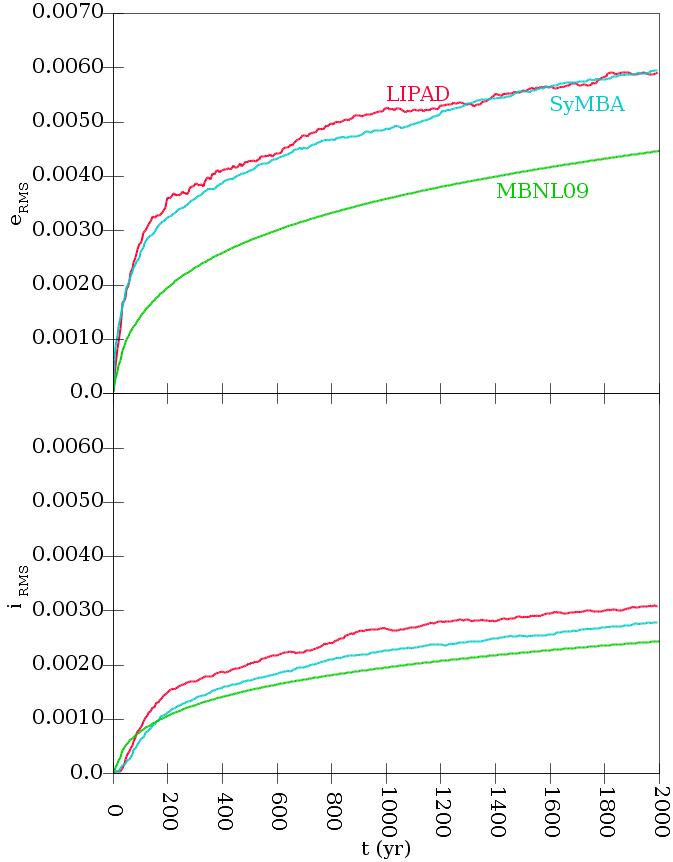}
   \caption{\footnotesize \label{fig:vsI} The temporal evolution of
     the RMS eccentricity (top) and inclination (bottom) of a system
     in the dispersion dominated regime.  In particular, the system
     contains 800 objects with $s = 540\,$km spread from 0.94 to
     $1.06\,$AU, and initial RMS eccentricity and inclination set to
     $3 \times 10^{-5}$.  The red, cyan, and green curves show the
     results from LIPAD, SyMBA, and MBNL09's code, respectively.  In
     the LIPAD run, the 800 particles were represented by 200
     tracers.}
\end{figure}

In the second simulation we test LIPAD's tracer viscous stirring
routines with a system that is both in the shear-dominated regime and
for which objects are isolated from one another (i.e$.$ objects are
separated enough from one another and dynamically cold enough that
their orbits do not cross).  As we explained above, MBNL09 argues that
it is important that these codes be able to handle such populations.
Here we study a narrow ring containing $0.001\,M_\oplus$ spread from
0.944 to 1.056 AU. This disk was populated with objects with $s =
93\,$km. The initial RMS eccentricity and inclination of these objects
was $7 \times 10^{-5}$.  For this system, the right hand side of
Eq.~\ref{eq:iso}, which defines whether a system is isolated, is
$0.033\,$AU.  Since this number is smaller than the width of the
annulus, which is $0.11\,$AU, then this system satisfies MBNL09's
isolation criterion (Eq.~\ref{eq:iso}).  The SyMBA run contains 610
particles, while we represented the same system with 152 particles in
LIPAD.

\begin{figure}[h!]
   \includegraphics[scale=0.5]{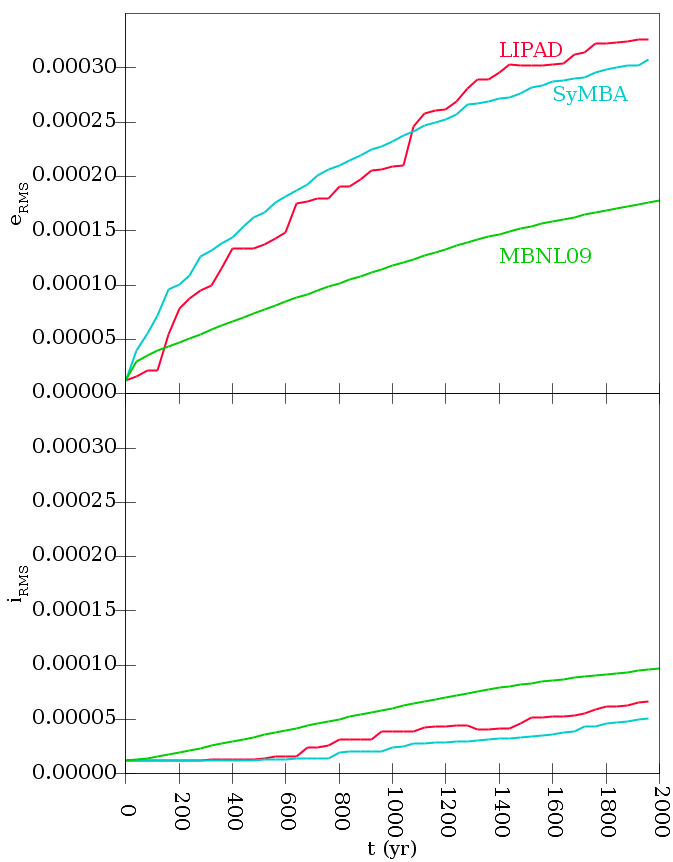}
   \caption{\footnotesize \label{fig:vsII} The temporal evolution of
     the RMS eccentricity (top) and inclination (bottom) of a system
     in the shear dominated regime.  In particular, the system
     contains 610 objects with $s = 93\,$km spread from 0.994 to
     $1.056\,$AU, and initial RMS eccentricity and inclination set to
     $7 \times 10^{-5}$.  This system also satisfies MBNL09's
     isolation criterion (Eq.~\ref{eq:iso}).  The red, cyan, and green
     curves show the results from LIPAD, SyMBA, and MBNL09's code,
     respectively.  In the LIPAD run, the 610 particles were
     represented by 152 tracers.}
\end{figure}

We show the temporal evolution of the RMS eccentricity and the RMS
inclination in our dispersion dominated runs in Figure~\ref{fig:vsI}
and our shear-dominated, isolated runs in Figure~\ref{fig:vsII}.  In
general, there is excellent agreement between LIPAD (shown in red) and
SyMBA (in cyan).  LIPAD does tend to excite inclinations slightly
faster than in a real $N$-body simulation, but this difference is
small.  In addition, LIPAD performs significantly better then the
analytic viscous stirring expressions used by MBNL09 and the other
Eulerian statistical codes.  Thus, we conclude that our new Monte Carlo
viscous stirring algorithm is the most accurate available for this
type of problem.

\subsection{Sub-Embryo Migration}
\label{ssec:Mig_test}

In \S{\ref{sssec:VSEm}} we described our methods for handling
planetesimal-driven migration for sub-embryos.
\citet{Kirsh:2009Icar:199:197} showed that an embryo must be at least
150 time more massive than the surrounding disk particles in order for
it to migrate.  Unfortunately, when a tracer is promoted to an embryo
it finds itself embedded in a disk of similar-mass objects, in spite
of the fact that these objects might be representing much smaller
planetesimals.  Thus, we were forced to develop a statistical way for
the small embryos (which we define as sub-embryos) to interact
directly with the planetesimals. Our method to accomplish this
involves performing a series of three body integrations, which include
the Sun, sub-embryo, and a single planetesimal, in order to determine
an average change in the energy of the sub-embryo (see
\S{\ref{sssec:VSEm}}).  This energy change is smoothly added to the
orbit of the sub-embryo via a fictitious acceleration applied to its
equation of motion.

\begin{figure}[h!]
   \includegraphics[scale=0.5]{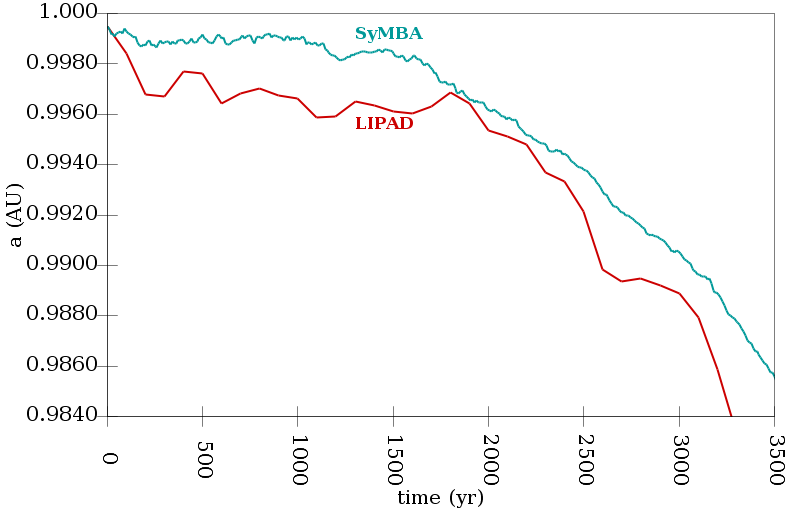}
   \caption{\footnotesize \label{fig:migr} The change in the
     semi-major axis of a $6.8\times 10^{-4}\,M_\oplus$ sub-embryo
     embedded in a $0.6\,M_\oplus$ disk spread from 0.915 to
     $1.085\,$AU.  The $N$-body simulation (cyan) contained 200,000
     disk particles, which were represented by 1000 tracers in LIPAD
     (red). This represents a migration of over $20\,r_H$.}
\end{figure}

We tested our methods using an experiment where we embed a $6.8\times
10^{-4}\,M_\oplus$ embryo in a $0.6\,M_\oplus$ disk consisting of
200,000 particles spread from 0.915 to $1.085\,$AU.  The embryo was
200 times more massive than the disk particles.  The cyan curve in
Figure~\ref{fig:migr} shows the migration of this embryo in an
$N$-body simulation using SyMBA.  After a period of 1500 years, the
embryo starts to migrate inward.  We performed the same simulation
with LIPAD where the 200,000 particles were replaced by 1000 tracers.
The ratio of the embryo mass to the tracer mass in this simulations
was 1.01.  The parameters of LIPAD were set so that the embryo was
considered to be a sub-embryo by the code. Perhaps, not surprisingly,
the LIPAD run was more noisy than the SYMBA run.  However, even under
these extreme conditions, the embryo's behavior was very similar in
the two simulations.

\subsection{Accretion with Velocity Evolution}
\label{ssec:KI}

Here we use the classic study of terrestrial planet accretion by
\citet{Kokubo:2000Icar:143:15} as a test of LIPAD.
\citet{Kokubo:2000Icar:143:15}'s goal was to perform the largest
$N$-body simulation of accretion to date, thereby giving them the
ability to start with the smallest possible planetesimals.  Indeed,
while the state-of-the-art simulations at the time started with a few
tens of objects, Kokubo \& Ida initially started with several thousand
bodies.  In order to start with even smaller initial objects, they
restricted their simulation to a narrow annulus.  In particular, they
constructed a distribution of objects initially in a ring that
extended from 0.99 to $1.01\,$AU and contained $0.3\,M_\oplus$ of
material. One of their runs contained 4000 particles with masses
between $10^{23}$ and $10^{24}$ gm; distributed in a power-law mass
distribution with a differential slope of $-2.6$. They followed the
system with a full $N$-body code. When particles collided they merged
--- there was no fragmentation.

   \begin{figure}[h!]
       \includegraphics[scale=0.5]{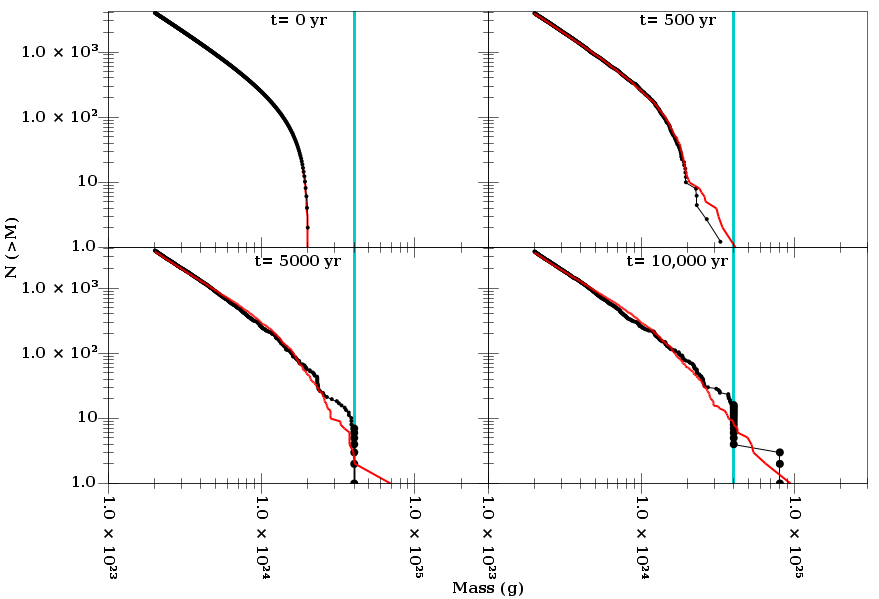}
       \caption{\footnotesize \label{fig:KI} Four snapshots of the
         evolution of the cumulative mass distribution of a system
         initially consisting of 4000 particles with masses between
         $10^{23}$ and $10^{24}$ gm for a total of $0.3\,M_\oplus$.
         This system was based on those in
         \citet{Kokubo:2000Icar:143:15}.  We represent this by 425
         tracers in LIPAD.  The back and red curves show the results
         from LIPAD and a direct $N$-body simulation done with SyMBA.
         The cyan line shows when a tracer is promoted to a
         sub-embryo. The full animation of this can be found at {\tt
           http://www.boulder.swri.edu/$\sim$hal/LIPAD.html}.}
   \end{figure}

We performed the same simulation with LIPAD, using 425 tracers
(approximately an order of magnitude fewer than in the $N$-body
calculation) {\color{MyChange} and $N_{\rm s-bin} = 20$}. Full
velocity evolution was included, but fragmentation was
disabled. Particles were promoted to embryos when they reached a mass
of $4\!\times\!10^{24}\,$ gm.  Figure~\ref{fig:KI} compares the
temporal evolution of the size-distribution from LIPAD (black) and
SyMBA (red).  The vertical cyan line shows the mass where tracers are
promoted to embryos.  Through this simulation all the embryos were
considered sub-embryos by the code.  Some explanation is required to
understand these results.

There are two phases of growth in the LIPAD simulation. Before 500
years, all the particles are tracers and thus the evolution of the
system is entirely done through our Monte Carlo routines.  Note that
there is excellent agreement during that time (see the top right panel
of Figure~\ref{fig:KI}, for example).  However, once an object becomes
a sub-embryo, it no longer interacts with the statistical part of the
code. As a result, it can only accrete tracers whole. In this
simulation tracers have $4\!\times\!10^{24}\,$gm and thus the
sub-embryos can only grow by accreting objects of this mass. This
limitation in the resolution of the growth rate near $m_{\rm tr}$
explains the pileup of objects at $4\!\times\!10^{24}$ and
$8\!\times\!10^{24}\,$gm.  This does not affect the overall growth
rates of the system, however.  Note that at the end of the simulation
at ${\color{MyChange}10},000$ years, the size-distributions from the
two codes nearly fall on top of one another.  Thus, as with the
collisional grinding, the resolution limitations of LIPAD do not seem
to affect the general behavior of the system.

\subsection{MBNL09's Final Test}

As a final test of their Eulerian particle-in-a-box planet accretion
code MBNL09 followed the evolution of $8.3\!\times\!10^8$ objects of
$4.8\!\times\!10^{18}\,$gm spread from 0.915 to $1.085\,$AU. These are
embedded in a gas disk with a mid-plane density at $1\,$AU of
$1.18\!\times\!10^{-9}\,$gm/cm$^3$, $\alpha=2.25$, and $z_s=0.05\,$AU
(see Eq.~\ref{eq:gden}).

The above problem represents an excellent opportunity to compare the
two types of algorithms because it is performed on such a narrow
annulus that planetesimals/planet migration (which the Eulerian codes
cannot handle) does not occur. In particular, we represented the
population of $8.3\!\times\!10^8$ planetesimals by 1000 tracers in
LIPAD {\color{MyChange} and set $N_{\rm s-bin} = 20$}. The smallest
object included in the collisional cascade is $10\,$m in size.  We
have turned on all of the velocity evolution routines with the
exception of the routine that allow sub-embryos to migrate.

   \begin{figure}[h!]
       \includegraphics[scale=0.5]{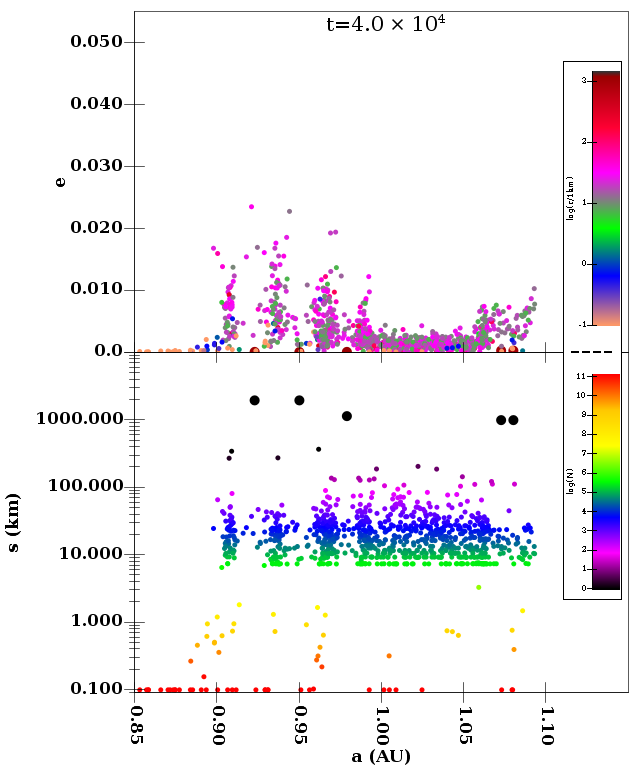}
       \caption{\footnotesize \label{fig:FT_part} A snapshot showing
         the tracers and embryos in our version of MBNL09's final test
         at $t = 40,000\,$yr.  We started with $8.3\!\times\!10^8$
         objects of $4.8\!\times\!10^{18}\,$gm spread from 0.915 to
         $1.085\,$AU, which we represented with 1000 tracers in
         LIPAD. Each dot represents a tracer (small dots) or an embryo
         (large dots; in this simulation all embryos are sub-embryos).
         The top panel presents an object's eccentricity ($e$) as a
         function of semi-major axis ($a$).  The color represents the
         radius of the object.  The bottom panel shows the objects
         radius ($s$) as a function of semi-major axis. In this case
         color shows $N_{\rm tr}$.  Note that the smallest object
         included in the size-distribution is $10\,$m in radius. The
         full animation can be found at {\tt
           http://www.boulder.swri.edu/$\sim$hal/LIPAD.html}.}
   \end{figure}

Figure~\ref{fig:FT_part} shows an animation of the evolution of the
system in LIPAD.  The top panel shows the eccentricity ($e$) of a
particle as a function of semi-major axes ($a$). The color of a dot
shows its size. Tracer particles and embryos are represented by small
and large dots, respectively. Note that all particles are tracers at
the beginning of the calculation. The bottom panel shows the radius
($s$) as a function of semi-major axes. Here the color indicates the
number of planetesimals the tracer actually represents. 

Recall that the main purpose of LIPAD was to be able to handle the
global redistribution of planetesimals due to the embryos.  Although
this particular test was designed to minimize this effect, we can
start to see the development of gaps around the embryos.  In addition,
although it is difficult to see in the figures, several of the embryos
have captured Trojan populations.

   \begin{figure}[h!]
       \includegraphics[scale=0.5]{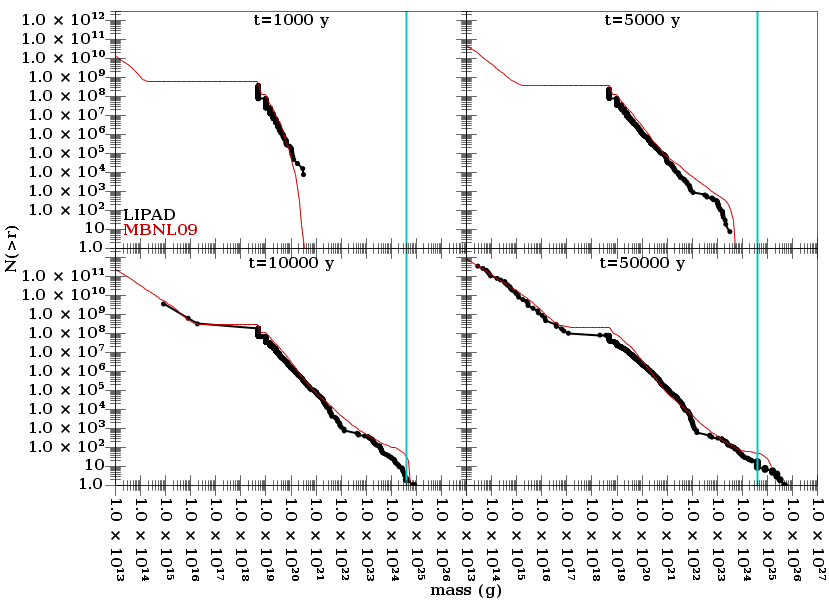}
       \caption{\footnotesize \label{fig:FT_SFD} Four snapshots of the
         evolution of the cumulative mass distribution during MBNL09's
         final test.  See Figure~\ref{fig:KI} for a detailed
         description of the figure.  The full animation can be found
         at {\tt http://www.boulder.swri.edu/$\sim$hal/LIPAD.html}.}
   \end{figure}

The remaining issue is to determine how well LIPAD and MBNL09 agree
with one another.  A comparison between the two simulations is shown
in Figure~\ref{fig:FT_SFD} (see caption for Figure~\ref{fig:KI} for a
description).  The match is reasonable. The largest discrepancies
occur in the first 5000 years. We attributed these differences to
differences in the velocity evolution. Recall that in our viscous
stirring tests presented in \S{\ref{ssec:VS_test}}, LIPAD did a better
job at the velocity evolution than MBNL09's code did. Thus, we believe
that LIPAD is probably correct in this case.

\section{Conclusions}
\label{sec:conl}

We presented the details of the first particle based (i.e$.$
Lagrangian) code that can follow the collisional/accretional/dynamical
evolution of a large number of km-sized planetesimals through the
entire growth process to become planets.  We refer to it as the {\it
  Lagrangian Integrator for Planetary Accretion and Dynamics} or {\it
  LIPAD}.  LIPAD is built on top of SyMBA, which is a symplectic
$N$-body integrator \citep{{Duncan:1998aj116:2067}}.  In order to
handle the very large number of planetesimals required by planet
formation simulations, we introduce four types of particles in LIPAD:
\begin{enumerate}
\item{} {\it Tracers:} These objects are intended to represent a large
  number of planetesimals on roughly the same orbit and size as one
  another.  Each tracer is characterized by three numbers: the
  physical radius $s$, the bulk density $\rho$, and the total mass of
  the disk particles represented by the tracer, $m_{\rm tr}$.  As a
  result, each tracer represents $N_{\rm tr}\!=\!m_{\rm tr}/{{4\over
      3}\pi\rho s^3}$ planetesimals.  They gravitationally interact
  with each other through Monte Carlo routines that include viscous
  stirring, dynamical fraction, and collisional damping
  (\S{\ref{ssec:tracer}}).  They are gravitationally stirred by the
  larger objects (i.e$.$ full- and sub-embryos, see immediately below)
  via the $N$-body routines.
\item{} {\it Full-Embryos:} These are the most massive objects in
  LIPAD.  They interact with all classes of particles through the
  direct summation of individual forces already present in the SyMBA
  code.  SyMBA routines also monitor whether physical collisions
  occur.  The algorithm that LIPAD uses to handle these collisions is
  described in \S{\ref{sssec:EmCol}}.
\item{} {\it Sub-Embryos:} These objects interact with full-embryos
  and each other through SyMBA routines.  However, the only dynamical
  effect that the tracers have on them is through dynamical friction
  and planetesimal-driven migration routines (\S{\ref{ssec:embryos}}).
  Collisions are handled in the same way as those of the full-embryos.
\item{} {\it Dust Tracers:} These are tracers that can no longer
  fragment.  The user can set the code so that these objects do not
  interact with the other tracers.  However, they always interact with
  the embryos via SyMBA's $N$-body routines.  The user also has the
  option to apply Poynting-Robertson drag.
\end{enumerate}

Perhaps LIPAD's greatest strength is that it can accurately model the
wholesale redistribution of planetesimals due to gravitational
interaction with the embryos, which has recently been shown to
significantly affect the growth rate of planetary embryos, themselves
(LTD10).  This redistribution controls growth in two ways.  First, it
can open gaps around the embryos thereby effectively stopping
accretion.  Additionally, the embryos can migrate as a result of
gravitational scattering of the near-by planetesimals, which can
enhance growth.  On the minus side, LIPAD struggles with being able to
accurately resolve the side-distribution of collisional tails.
However, we show that this does not affect embryo growth rates.

We have carefully verified and tested LIPAD.  In \S{\ref{sec:tests}},
we present experiments that independently exercise all of LIPAD's
abilities.  We find that it out performs Eulerian statistical
algorithms previously used to study this problem.  Of particular note,
LIPAD's viscous stirring routines are particularly accurate.

Our Lagrangian approach has an advantage over most previous attempts
to study planet formation because, rather than using analytical
expressions to estimate the global evolution of the system, our code
mimics the important micro-physics (i$.$e., local accelerations and
individual collisions) and lets the global system evolve naturally.
Therefore, there are fewer assumptions made, and the interactions
between different mechanisms are handled more realistically.

{\color{MyChange} LIPAD has many free parameters.  In addition, we
  have not discussed any issues concerning the convergence of the
  code.  Of course, each problem is different and so a user is going
  to be required to investigate these issues on their own.  Thus, we
  decided that the best approach is to present an illustrative example
  of a production run we are currently undertaking with LIPAD.  In
  particular, we are doing a series of simulations of terrestrial
  planet formation in the region between 0.7 and $1.5\,$AU.  This
  region of the planetary system is populated with $2.9\,M_\oplus$ of
  planetesimals whose initial radii varied from 10 to $50\,$km
  depending on the run.  The surface density of the planetesimals
  initially scale as $r^{-1.5}$. We represent these planetesimals with
  $12,000$ tracers of $1.4 \times 10^{24}\,$g (roughly 50\% more
  massive than Ceres).  This implies that tracers transition to
  sub-embryos when they have a radius of only $\sim\!450\,$km.  The
  transition from sub-embryo to embryo is set to $2.8 \times
  10^{26}\,$g, roughly 200 times a tracer mass.  We use $N_{\rm
    s-bin}=30$ spanning sizes from 1 to $450\,$km, and have 166
  annular bins stretching from 0.5 to $60\,$AU. We set $s_{\rm dust} =
  30\,\mu$ and adopt BA99's values for high velocity rock in the
  calculations of $Q^*_D$. We embed this material in a minimum mass
  gas disk \citep{Hayashi:1985prpl.conf:1100}, which we assume has an
  opacity of 2\% the ISM value when in the atmospheres of the embryos.
  Type~I eccentricity damping is included with $c_e=1$, but we disable
  Type~I migration by setting $c_a=0$.  The gas disk decays with a
  lifetime of $2\,$My.  The timestep for the $N$-body code is set to
  0.025 years, while the one for the statistical code is 3 times
  longer. We will present an analysis of these calculations in an
  upcoming paper.  }

\acknowledgments We would like to thank Glen Stewart and Alessandro
Morbidelli for useful discussions{\color{MyChange}, and to John
  Chambers who acted as referee on this manuscript}.  This work as
been directly supported by a grant from the National Science
Foundation. This project was also supported by the Center for Lunar
Origin and Evolution (CLOE) of NASA's Lunar Science Institute (Grant
Number NNA09DB32A).  HFL is also grateful for funding from NASA's
Origins, and OPR programs.  MJD acknowledges the continuing financial
support of NSERC, Canada.

\clearpage
\bibliographystyle{elsarticle-harv}
\bibliography{allrefs2}
\clearpage

\end{document}